\documentclass[]{aastex631}
\usepackage{amssymb}
\usepackage{amsmath}
\usepackage{appendix}



\shortauthors{Chen $\&$ Dai}
\shorttitle{Jets in AGN disk}

\begin{document}
	
	\title{Observational Properties of Thermal Emission from Relativistic Jets Embedded in AGN Disks}
	
	\author[0000-0001-8955-0452]{Ken Chen}
	
	\affiliation{School of Astronomy and Space Science, Nanjing University, 
		Nanjing 210023, China}
	
	\author[0000-0002-7835-8585]{Zi-Gao Dai}
	
	\affiliation{Department of Astronomy, School of Physical Sciences, 
		University of Science and Technology of China, Hefei 230026, China; daizg@ustc.edu.cn}
	
\begin{abstract}
Relativistic jets can be produced within the accretion disk of an active galactic nucleus (AGN), leading to distinct thermal emission as they propagate through a dense disk environment. In this paper, we present a comprehensive study of dynamical evolution of jets embedded in an AGN disk and their associated observational properties, focusing on scenarios in which jets either successfully break out of the disk or become choked. By modeling the jet-cocoon system propagation, we calculate the thermal emission contributions from the jet-head shock breakout, disk cocoon, and jet cocoon components. Our results reveal that soft X-ray flares are the most prominent observable signatures, with duration ranging from $O(10^2)\,\text{s}$ to $O(10^5)\,\text{s}$, 
occasionally exhibiting double-peaked light curves, whereas UV/optical flares are detectable only for powerful jets, persisting for several days to tens of days. This thermal emission serves as a critical electromagnetic counterpart to jet-producing events and provide insights into jet dynamics and AGN disk properties. Our findings highlight the importance of multi-wavelength follow-up observations to establish a diagnostic paradigm for candidate electromagnetic counterpart identification to AGN-embedded events and to distinguish thermal flares 
from AGN background variability.
\end{abstract}

\keywords{Active galactic nuclei (16); Accretion (14); Relativistic jets (1390)}

\section{Introduction}
The accretion disk of an active galactic nucleus (AGN) serves as 
a cosmic laboratory for extreme astrophysical phenomena, 
many of which are associated with the production of relativistic 
jets. The AGN disk is predicted to harbor stars and compact 
objects \citep[e.g.][]{Syer91, Goodman03, Mckernan12,
Fabj20, Jermyn21, Fabj25}, In such an environment, the core-collapse 
of massive stars can generate jets powering long gamma-ray burst 
(GRB) \citep{Woosley93, Woosley06}, while compact objects 
undergoing hyper-Eddington accretion processes are capable 
of launching jets \citep[e.g.][]{Tagawa23a}. Meanwhile, the AGN 
disk provides a plausible formation channel for gravitational wave (GW) 
events \citep[e.g.][]{Abbott20, Tagawa20, Mckernan20a, Mckernan20b, 
Perna21b, Ishibashi24, Delfavero24, Rowan24}. Binary neutron star 
mergers or neutron star-black hole mergers embedded in the AGN disk 
can produce jets that power short GRBs \citep[e.g.][]{Abbott17a, 
Abbott17b}, whereas the remnants of binary black hole mergers 
can launch jets via accreting the ambient dense gas \citep[e.g][]
{Tagawa23b, Tagawa24, Rodrguez24, ChenK24}. Tidal disruption 
events involving stellar-mass black holes in the AGN disk 
\citep{Yang22} could also produce jets similar to those observed 
in jetted tidal disruption events \citep[e.g.][]{Dai18}. 
Investigating the emission generated by 
jets embedded in AGN disks facilitates the prediction of 
electromagnetic (EM) counterparts of jet-producing systems, 
thereby enabling a better understanding of their properties.
Simultaneously, jet emissions can provide valuable insights 
into the nature of the AGN disk.

Recently, several studies have explored the observational
properties of two specific types of jets within AGN disks: 
GRB jets and GRB-less jets produced in binary 
black hole merger systems. For GRB jets, 
significant attention has been focused on the production 
and propagation of non-thermal radiation. Both the intrinsic 
properties of prompt emissions \citep[e.g.][]{Perna21a,
Lazzati22, Kang25} and afterglows \citep[e.g.][]{Wang22, 
Kathirgamaraju24, Huang24, Kang25} of embedded GRB jets 
differ substantially from typical cases \citep{ZhangB}. 
These non-thermal photons undergo frequent diffusion 
\citep[e.g.][]{Perna21a, Wang22} and absorption 
\citep[e.g.][]{Ray23} before escaping the disk photosphere, 
which further modifies the observed emissions. Additionally, 
\cite{Zhang24} examined the dynamics of GRB jets within
AGN disks and found that these jets are probably choked by
the dense environment. 
On the other hand, the potential observation of jets from 
the remnants of binary black hole mergers is explored
through analyzing their interaction with the AGN disk 
\citep{Tagawa23b, Tagawa24, ChenK24, Rodrguez24}. The
resulting emissions originate from the jet-head successfully
breaking out of the disk \citep{Tagawa23b, ChenK24} and 
the subsequent expansion of shocked jet and disk materials
\citep{Tagawa24, ChenK24, Rodrguez24}.
However, these studies have been confined to specific
astrophysical events and each demonstrates certain limitations.
For instance, some have oversimplified the interaction between
the jet and the AGN disk, thereby neglecting the formation of
the cocoon structure; some have failed to consider photon 
trapping within the opaque jet-cocoon system; while others have
omitted a comprehensive evaluation of radiation contributions 
from all material components. Therefore, a systematic 
investigation that simultaneously address both jet propagation 
and observed radiation production becomes imperative and valuable.

In this paper, we investigate the observational properties of 
a relativistic jet with a power and a duration launched at
different radius within various AGN disks. We calculate the 
entire propagation process of the jet until it breaks out of 
the AGN disk, focusing on the thermal emissions from the opaque 
jet-cocoon system generated by the interaction between the jet
and the AGN disk. These emissions either serve as the first 
light of a successfully breaking-out jet or represent the
sole signal of a choked jet.
The paper is organized as follows. In Section \ref{section2},
we describe the dynamics of jet propagation and investigate
the properties of the jet-cocoon system, including the 
jet-choking criterion. In Section \ref{section3}, we present
a comprehensive analysis of the thermal emission 
from all components of the system, namely the jet-head 
shock breakout emission, the disk-cocoon emission, and 
the jet-cocoon emission. Additionally, we examine the emission 
mechanism in the jet-choking scenario. The observational 
properties of radiation produced by the embedded jet are 
outlined in Section \ref{section4}, with particular emphasis 
on the detectability of these thermal emissions in Section 
\ref{section4-2}. We summarize our main results and provide 
discussion in Section \ref{section5}.
Symbols $G$ and $c$ in this paper represents the gravitational 
constant and the speed of light, respectively.

\section{Jet Propagation within AGN disk}
\label{section2}
\subsection{AGN Disk Structure}
To investigate the propagation of jets in an AGN disk, it is essential to 
determine the vertical density profile of the disk gas. An accretion disk is commonly understood 
to maintain a vertical hydrostatic equilibrium between its pressure and gas self-gravity \citep{Kato08}. 
Adopting a one-zone approximation to simplify the vertical structure, where the density is assumed
to be uniform with $\rho_{\text{uni}}\left(z\right)=\rho_{0} $ and truncated at the disk surface of
a scale height $H_{\text{d}}$, various models have been developed to characterize 
the structures of AGN disk \citep[e.g.][]{Sirko03, Thompson05, Gilbaum22, ChenYX24a, Daria24, Marguerite25}.
Two more plausible vertical density profiles can be obtained by incorporating the specific
pressure properties, one is the gas-pressure-dominated isothermal density profile \citep{Kato08}
\begin{equation}
\rho_{\text{iso}}\left(z\right)=\rho_{0}\exp\left[-\frac{z^2}{2H_{\text{d}}^2}\right],
\end{equation} 
and another is the radiation-pressure-dominated polytropic density profile \citep{Kato08, Grishin21}
\begin{equation}
\rho_{\text{poly}}\left(z\right)=\rho_{0}\left(1-\frac{z^2}{6H_{\text{d}}^2}\right)^3.
\end{equation} 
Both profiles indicate that the density is roughly uniform within $H_{\text{d}}$ 
and drop sharply beyond $H_{\text{d}}$, suggesting a more diffuse structure compared 
to $\rho_{\text{uni}}$.

\begin{figure*}
	\begin{center}
		\includegraphics[width=0.4\textwidth]{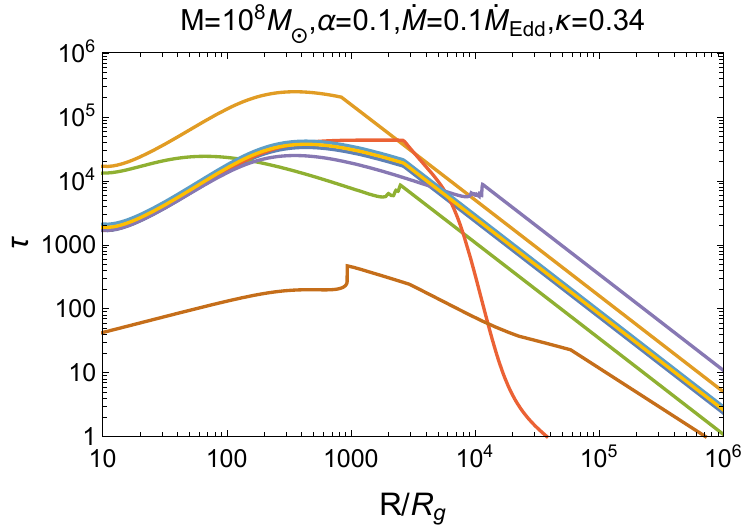}
		\quad
		\includegraphics[width=0.4\textwidth]{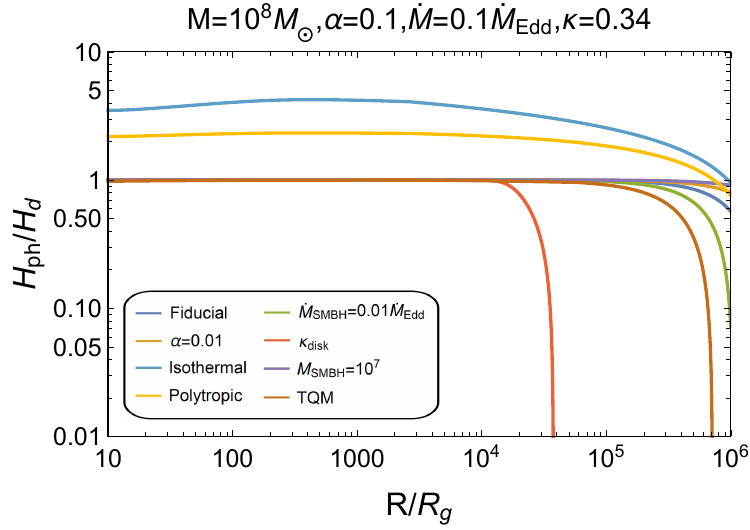}
	\end{center}
	\caption{Vertical optical depth and photosphere height measured from the mid-plane of AGN disk. 
		$M$ is the mass of supermassive black hole (SMBH) 
		in units of $M_\odot$, $\alpha$ is the viscosity parameter \citep{SS73}, and 
		$R$ is the radius of AGN disk on scale of gravitational radius $R_{\text{g}}=GM/c^2$.
		The mass inflow rate of AGN disk is in units of the Eddington limit accretion rate, 
		$\dot{M}_{\text{Edd}}=L_{\text{Edd}}/\eta c^2$, where $\eta=0.1$ and
		$L_{\text{Edd}}=4\pi G M m_{\text{p}}c/\sigma_{\text{T}}=1.26\times10^{46}(M/10^8M_\odot)\text{\,erg\,s}^{-1}$, 
		$m_{\text{p}}$ is the proton mass and $\sigma_{\text{T}}$ is the Thomson cross section. 
	    The fiducial disk structure adopts \cite{Sirko03} model with parameters
	    $M=10^{8}\text{M}_{\odot}$, $\dot{M}=0.1 \dot{M}_{\text{Edd}}$, $\alpha=0.1$, and opacity 
	    $\kappa=0.34\text{\,cm}^{2}\text{\,g}^{-1}$. Different disk parameters are shown in the legend, where 
	    $\kappa_{\text{disk}}$ represents the opacity as a function of density and temperature 
	    which is set as an approximate form shown in \cite{Yang19b}; TQM denotes that the \cite{Thompson05}
        disk model is adopted.}
	\label{Fig:tau}
\end{figure*}

As the energy source for a supermassive black hole at the galactic center, both the mass of an AGN disk 
and its mass inflow rate must be sufficiently high to power a luminous active galactic nucleus.
Although the AGN disk is generally geometrically thin with $H_{\text{d}}/R\ll1$, where $R$ is 
the disk radius, its actual vertical thickness combined with high density can result in an 
opaque gas environment. The vertical optical depth of the disk is given by
\begin{equation}
\tau_{\text{d}}\left(z\right)=\int_{z}^{z_{\text{edge}}} \kappa \rho\left(z'\right) \text{d}z',
\end{equation}  
where the disk surface $z_{\text{edge}}$ is set as $H_{\text{d}}$, $\sqrt{6}H_{\text{d}}$, and infinity for 
the uniform disk, polytropic disk, and isothermal disk, respectively. As shown in Figure \ref{Fig:tau},
the inner region of the AGN disk is extremely optically thick with $\tau_{\text{d}}(z=0)\gg1$, and the 
photosphere location is close to the disk surface. Both polytropic and isothermal disk exhibit a higher
photosphere due to their more extended gas distribution, yet they maintain a similar total optical depth 
to the uniform profile because the density remains low beyond $H_{\text{d}}$.

Since we investigate the thermal emissions from an initially opaque jet system, and
various astrophysical events that generate jets typically occur in the inner region of the AGN
disk, such as binary stellar-mass black hole mergers 
\citep{Yang19a, Tagawa20, Mckernan20a, Rowan24, Xue25}, black
hole-neutron star or binary neutron star mergers \citep{Mckernan20b, Perna21b}, and stellar-mass black hole
hyper-Eddington accretion \citep{Tagawa23a}, we focus on disk regions where $R\leqslant10^5R_{\text{g}}$. 
The \cite{Sirko03} model is employed, which well describes these regions, and we set disk parameters 
as $\alpha=0.1$ and $\dot{M}=0.1 \dot{M}_{\text{Edd}}$ below.

\subsection{Jet Dynamics}
\begin{figure*}
	\begin{center}
		\includegraphics[width=0.3\textwidth]{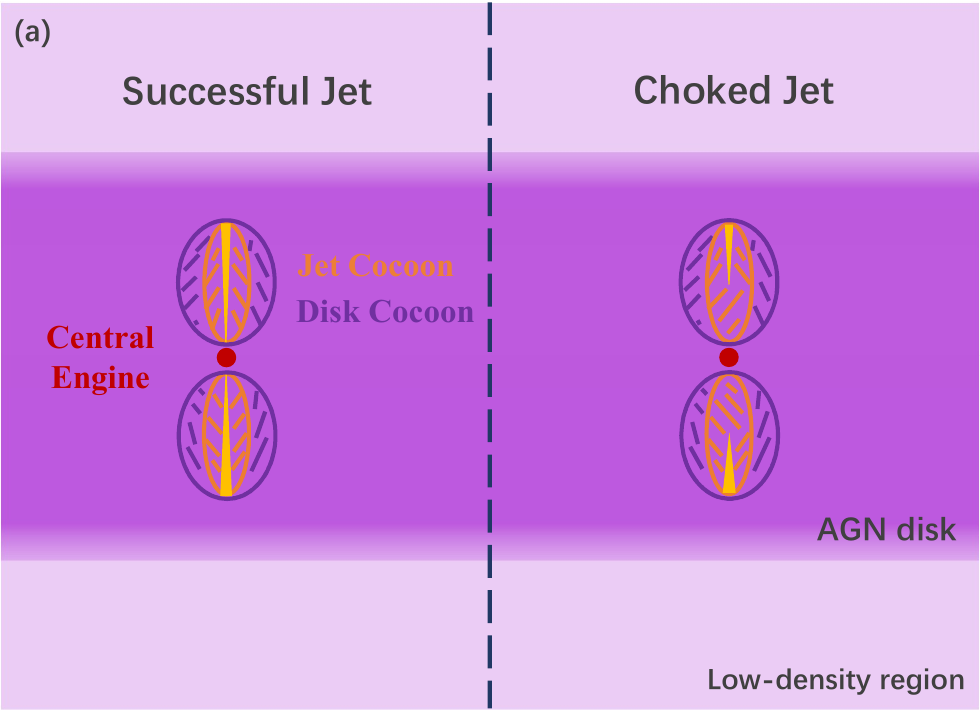}
		\quad
		\includegraphics[width=0.3\textwidth]{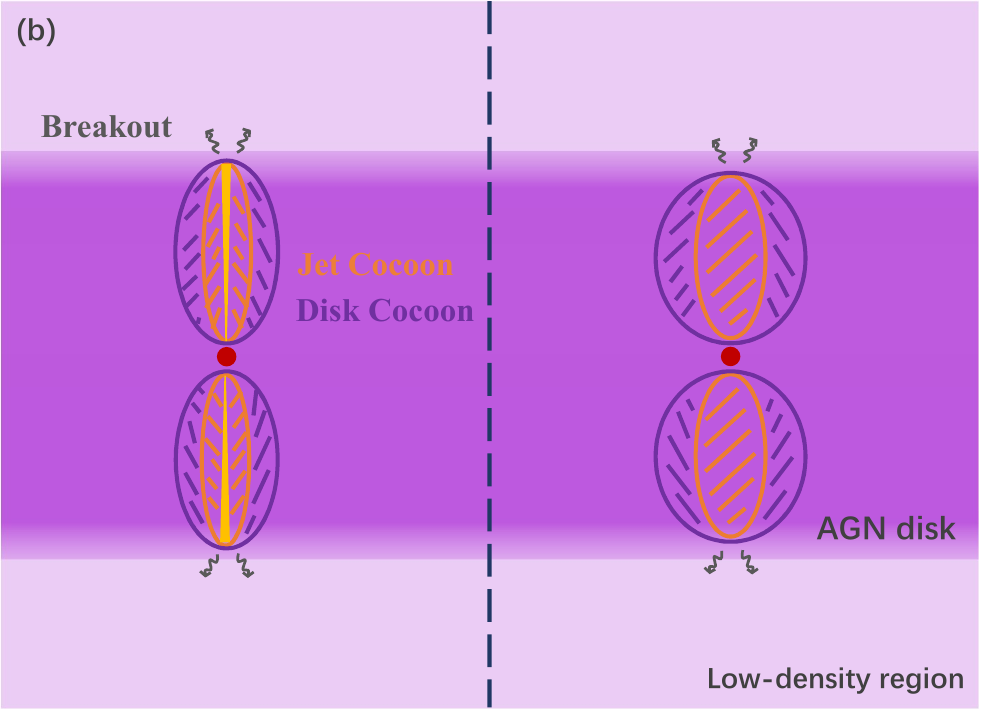}
		\quad
		\includegraphics[width=0.3\textwidth]{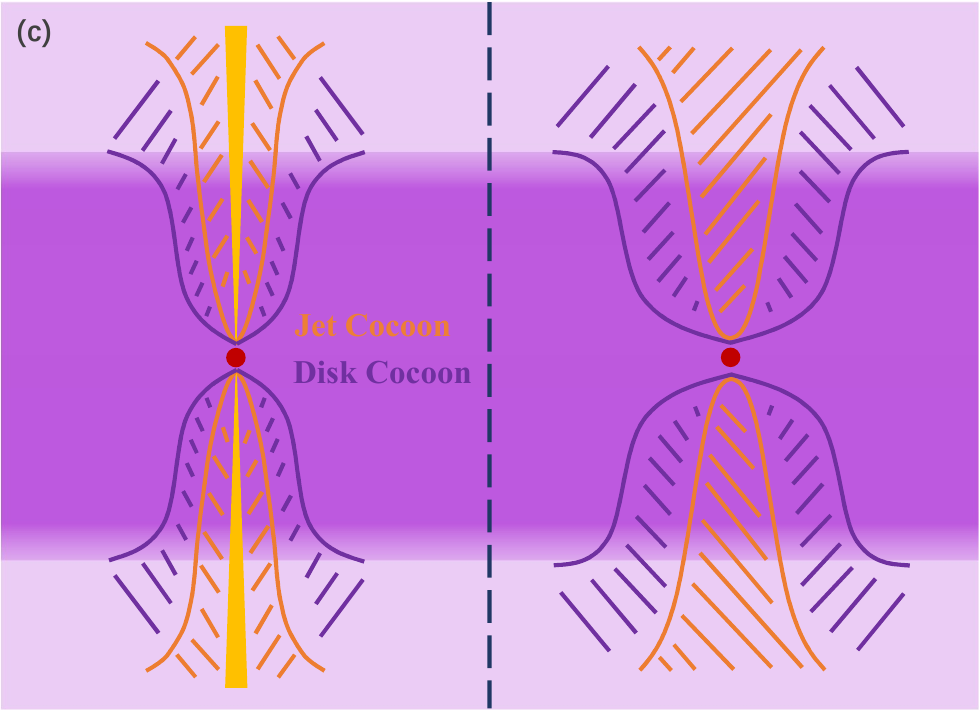}
	\end{center}
	\caption{Schematic description of the jet propagation in the AGN disk. (a) Depending on the
	duration of the central engine, a jet can either successfully penetrate the disk or become 
	choked. Prior to engine cut-off and before the jet's tail catching up with its head, both successful 
	and choked jet system exhibit similar structure: a jet-body enveloped by a two-component cocoon.
    (b) For the successful jet, the ram pressure exerted on the jet head facilitates rapid forward
    propagation, which in turn maintains the slender shape of the jet-cocoon system. In contrast,
    once the jet is choked and dissipates, the cocoon expands both longitudinally and laterally
    driven by its internal pressure, resulting in a wider structure. At the breakout, photons can diffuse out 
    from either the jet-head or the cocoon-head to produce emission. 
    (c) No matter whether the jet is choked or not, the two-component cocoon expands ahead of AGN disk and
    continuously emits thermal radiation.}	
	\label{Fig:sch}
\end{figure*}

Launched from the central engine and propagating within the AGN disk, 
a jet interacts with its dense ambient gas at its head. 
The extremely optically thick environment inhibits photon escape, causing
the shocked disk material and jet material to flow laterally from the jet-head and 
form a pressurized cocoon that envelops the jet body. A schematic diagram is depicted in 
Figure \ref{Fig:sch}. The structure and evolution of the 
jet-cocoon system have been extensively investigated in the literature
\citep[e.g.][]{Begelman89, Matzner03, Bromberg11, Harrison18, 
	Gottlieb20, Gottlieb21, Hamidani21, Urrutia23, Suzuki24},
with studies demonstrating that self-consistent analytic models are in good agreement with 
numerical simulations \citep{Bromberg11}. In line with our previous work \cite{ChenK24}, 
we employ the descriptions in \cite{Bromberg11} to calculate the jet-cocoon dynamics.

The jet-cocoon system is characterized by three key parameters: the movement distance of the jet head 
$z_{\text{h}}$, the internal energy stored in the cocoon $E_{\text{c}}$, and the lateral radius of 
the cocoon $r_{\text{c}}$, which evolve over time as 
\begin{gather}\label{eq:dt}
\frac{\text{d} z_{\text{h}}}{\text{d}t}=\beta_{\text{h}}c,\\
\frac{\text{d} E_{\text{c}}}{\text{d}t}=
\eta_{\text{h}} L_{\text{j}}\left(1-\beta_{\text{h}}\right),\\
\frac{\text{d} r_{\text{c}}}{\text{d}t}=\beta_{\text{c}}c,
\end{gather}
where $L_{\text{j}}$ is the jet power. The jet-head velocity is determined by the 
ram pressure balance between jet and disk medium 
\citep{ Matzner03}, which can be approximated as
\begin{equation}
\beta_{\text{h}}=\frac{\beta_{\text{j}}}{1+\tilde L^{-\frac{1}{2}}},
\end{equation}
where $\beta_{\text{j}}\sim1$ is the jet velocity, and $\tilde L$ is a dimensionless parameter 
\begin{equation}
\tilde L=\frac{L_{\text{j}}}{\Sigma_{\text{h}}\rho(z) c^3}.
\end{equation}
$\eta_{\text{h}}$ represents the fraction of energy transferred from the jet head to the cocoon,
depending on the causal connection of the shocked material and determined by the Lorentz factor
$\Gamma_{\text{h}}$ and the opening angle $\theta_{\text{h}}$ of the jet head, which is
\begin{equation}
\eta_{\text{h}}=\min[\frac{2}{\Gamma_{\text{h}}\theta_{\text{h}}},1].
\end{equation}
Squeezed by the highly pressurized cocoon, the jet would be collimated from a conical to a 
cylindrical structure. The resulting jet cross-section is
\begin{equation}
\Sigma_{\text{h}}=\pi\theta_{\text{j}}^2\min[z_{\text{h}}^2,
\frac{3L_{\text{j}}z_{\text{h}}r_{\text{c}}^2}{4cE_{\text{c}}}],
\end{equation}
where $\theta_{\text{j}}$ is the initial opening angle of the jet, 
the two terms represent the 
undisturbed case and the collimated case, respectively. The lateral expansion velocity of the
cocoon can be estimated as $\beta_{\text{c}}=\sqrt{P_{\text{c}}/\overline\rho(z_{\text{h}})c^2}$. 
Given that the pressure of a radiation-dominated cocoon is $P_{\text{c}}=E_{\text{c}}/3V_{\text{c}}$, 
and its volume can be approximated as $V_{\text{c}}=\pi z_{\text{h}}r_{\text{c}}^2$, the expression for
the velocity becomes
\begin{equation}\label{eq:bc}
\beta_{\text{c}}=\left[\frac{E_{\text{c}}}
{3\pi\overline \rho(z_{\text{h}})c^2z_{\text{h}}r_{\text{c}}^2}\right]^{\frac{1}{2}},
\end{equation}
where $\overline \rho(z_{\text{h}})$ is the averaged density of the cocoon at $z_{\text{h}}$. 
Combining Equations (\ref{eq:dt})-(\ref{eq:bc}), 
the comprehensive dynamical evolution of the
jet-cocoon system can be calculated self-consistently. For simplicity, the jet is assumed to 
be always launched at the mid-plane of AGN disk and propagates perpendicular to the disk plane,
with initial parameters $\Gamma_{\text{j}}=100$ and $\theta_{\text{j}}=0.17$.

\begin{figure*}
	\begin{center}
		\includegraphics[width=0.32\textwidth]{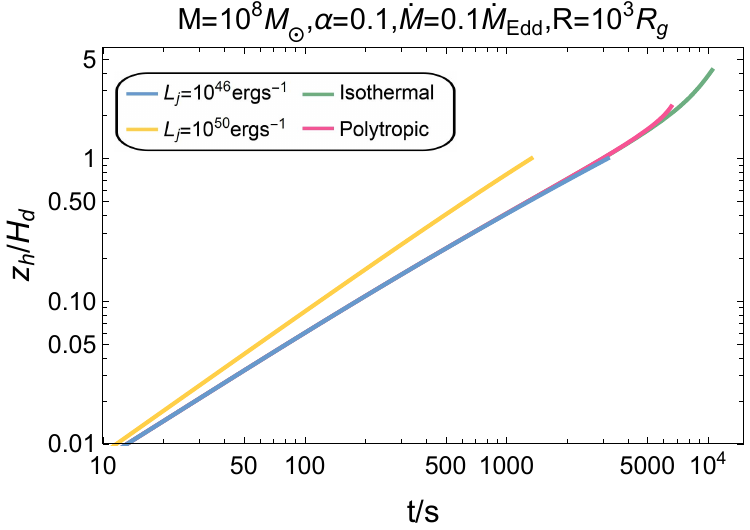}
		\includegraphics[width=0.32\textwidth]{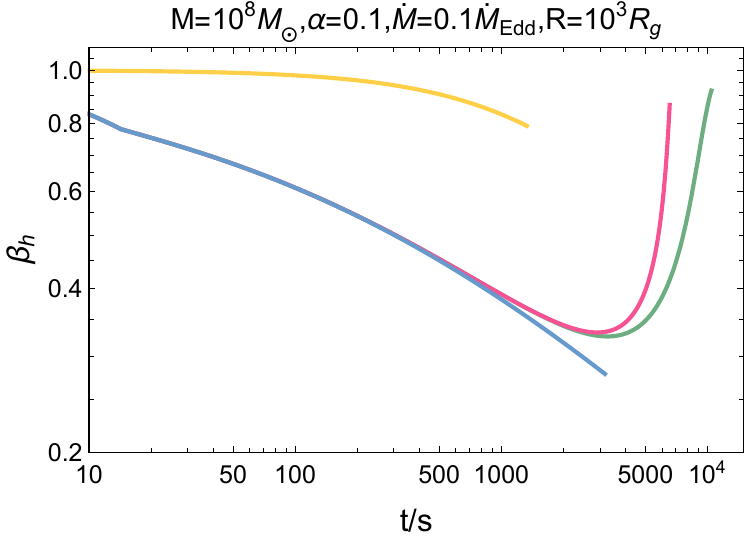}
		\includegraphics[width=0.32\textwidth]{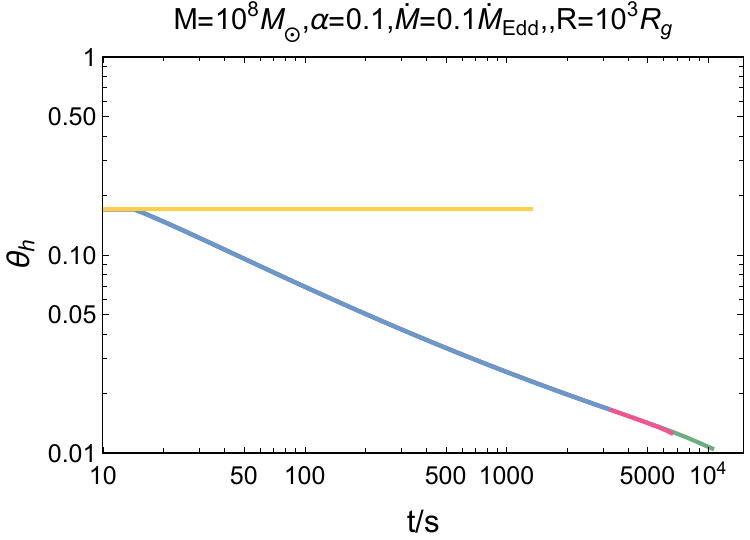}
		\includegraphics[width=0.32\textwidth]{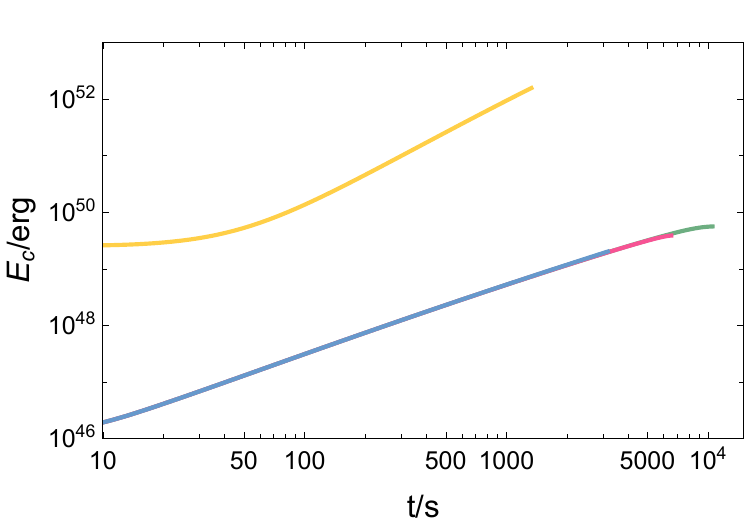}
		\includegraphics[width=0.32\textwidth]{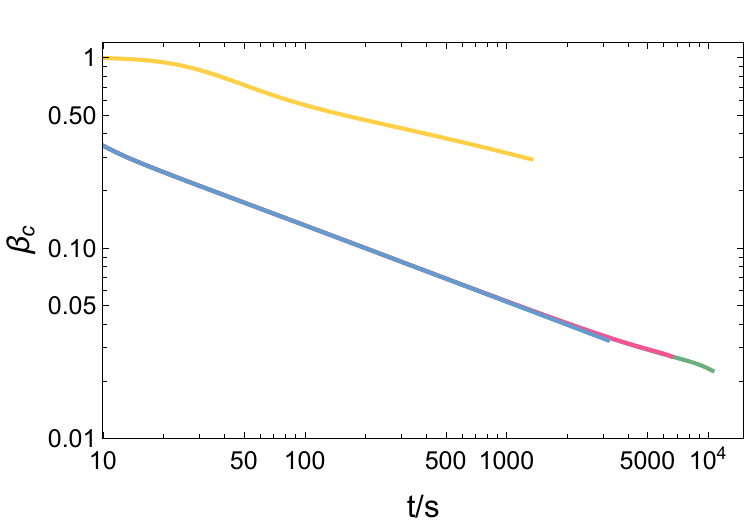}
		\includegraphics[width=0.32\textwidth]{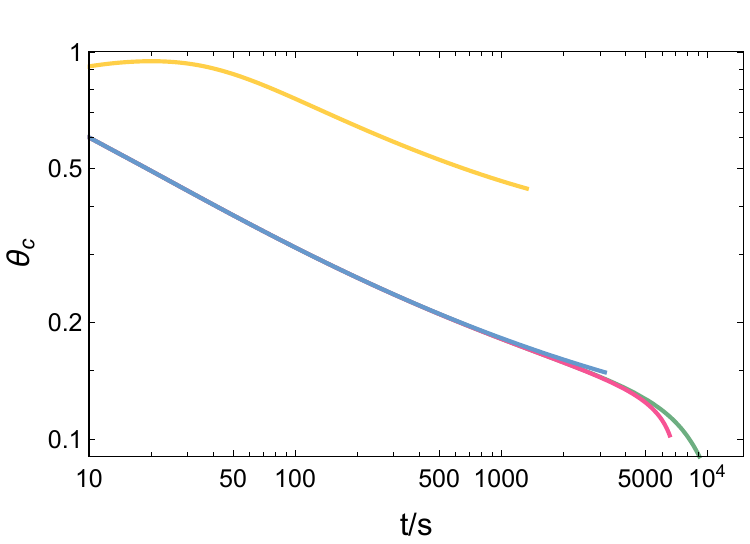}
	\end{center}
	\caption{Evolution of system parameters in illustrative jet-cocoon systems. 
	The blue, yellow, green, and magenta lines represent the propagation 
	dynamics of a $10^{46}\text{\,erg\,s}^{-1}$ jet embedded at the mid-plane 
	of an AGN disk with a uniform density profile, as well as jet with higher 
	power of $10^{50}\text{\,erg\,s}^{-1}$ or different density profiles. 
	The endpoint of each line indicates the time of jet breakout.}
	\label{Fig:dyn}
\end{figure*}

Four illustrative examples for the evolution of a jet-cocoon system are shown in Figure \ref{Fig:dyn}.  
The $L_{\text{j}}=10^{46}\text{\,erg\,s}^{-1}$ case clearly demonstrates the essential
characteristics of the jet-cocoon system. The jet is significantly decelerated 
and collimated during its propagation. Additionally, the vertical motion of the
jet-head is much more rapid than the lateral expansion of the cocoon, resulting
in an elongated system. When the jet is extremely strong, for instance, with 
$L_{\text{j}}=10^{50}\text{\,erg\,s}^{-1}$, the ambient gas cannot effectively
decelerate the jet, and consequently, the velocity of the jet-head is higher.
Despite the significant internal pressure, the cocoon remains too weak
to collimate the jet, leading to a nearly constant jet opening angle throughout 
its propagation. Due to the similarity between $\rho_{\text{uni}}$, $\rho_{\text{iso}}$,
and $\rho_{\text{poly}}$ within $H_{\text{d}}$, the dynamics of jet is nearly
indistinguishable when its head is deeply embedded within the AGN disk for these three
vertical density profile. Since the isothermal and polytropic disk are more extended,
the jet would continuously propagate within the AGN disk ahead of $H_{\text{d}}$. 
Moreover, as the ambient density decreases sharply, the tenuous gas inefficiently 
brakes the jet, allowing it to reaccelerate. For the specific GRB jets, \cite{Zhang24}
demonstrated similar dynamical properties of jet propagation to those presented in our 
study.

\subsection{Jet Choking}

\begin{figure*}
	\begin{center}
		\includegraphics[width=0.315\textwidth]{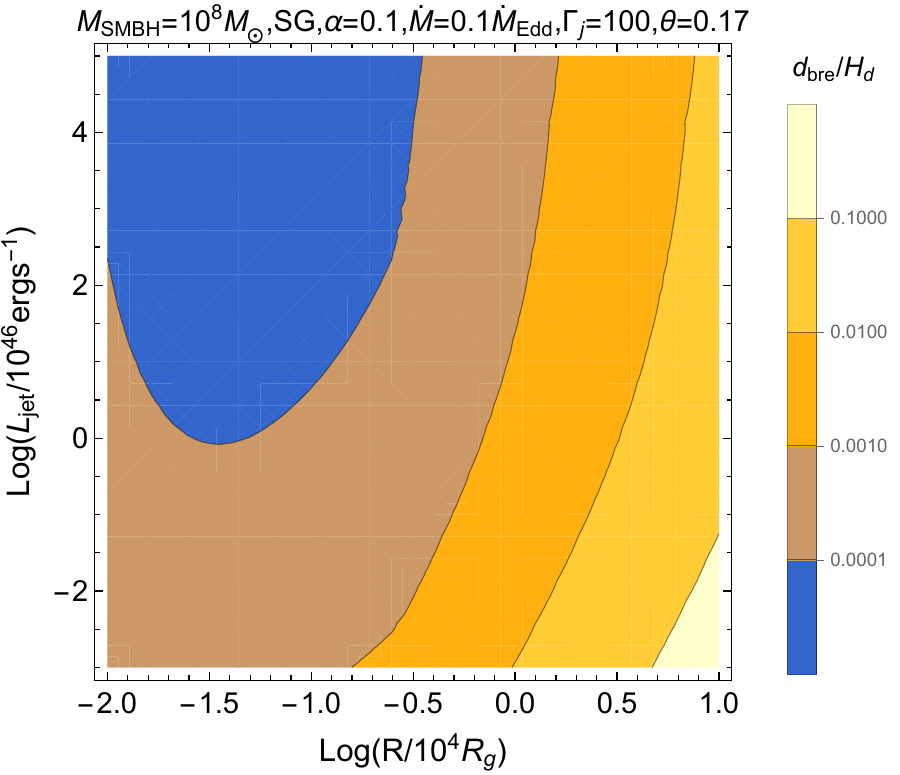}
		\quad
		\includegraphics[width=0.31\textwidth]{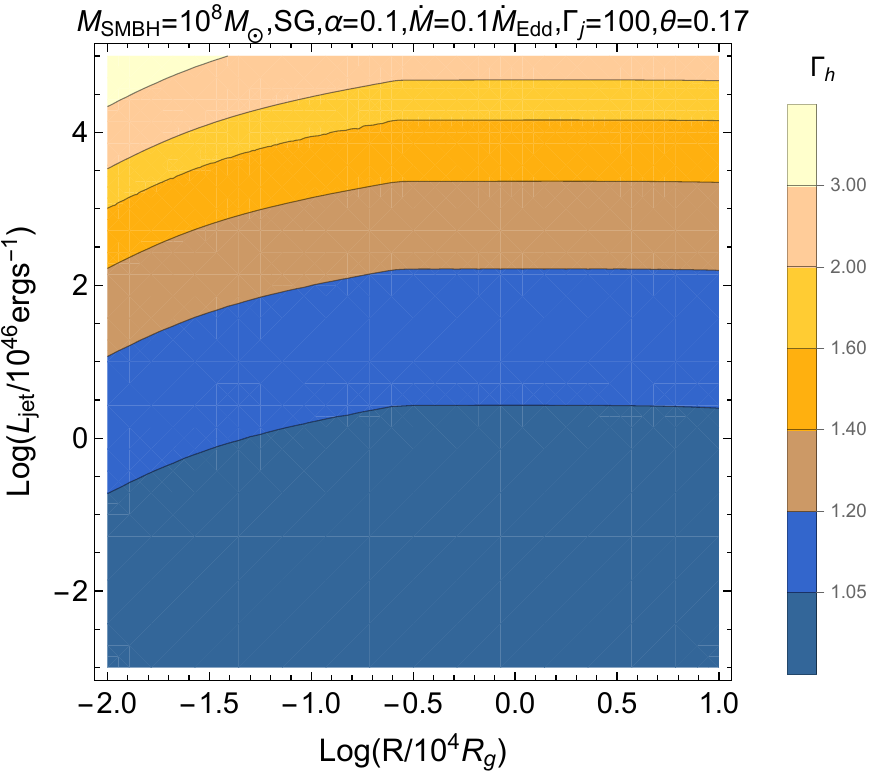}
		\quad
		\includegraphics[width=0.31\textwidth]{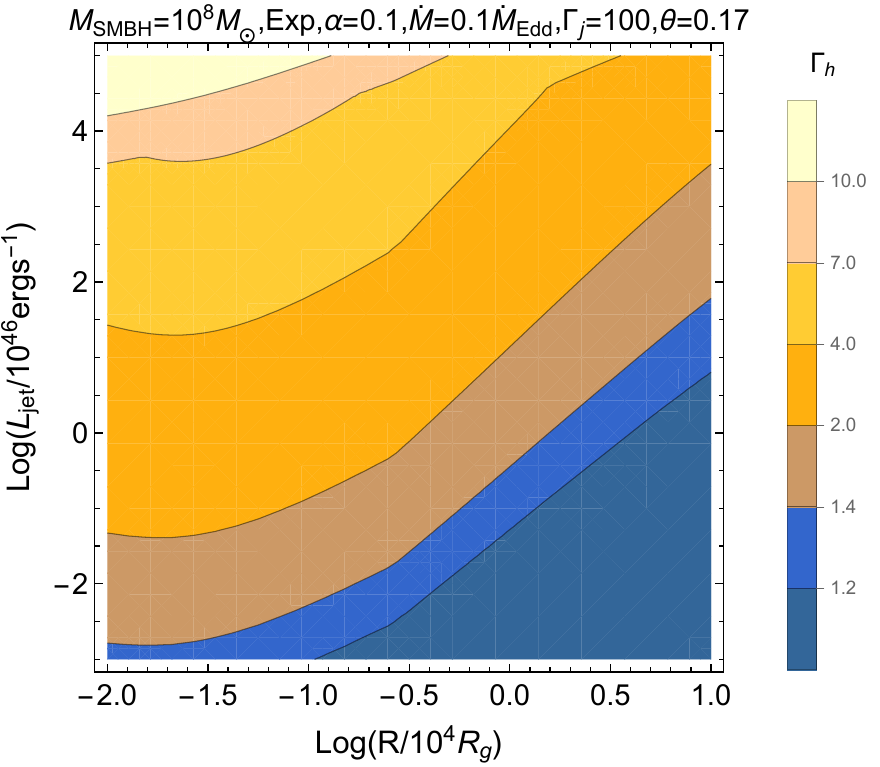}
	\end{center}
	\caption{Properties of the jet head at shock breakout. The left and middle panels 
		respectively depict the width of the shock breakout shell 
		(in units of the disk height $H_{\text{d}}$) and the Lorentz factor 
		of the jet head at breakout, assuming a uniform disk density profile. The right 
		panel shows the Lorentz factor of the jet head at breakout under an isothermal 
		disk density profile.}	
	\label{Fig:bre}
\end{figure*}

\begin{figure*}
	\begin{center}
		\includegraphics[width=0.31\textwidth]{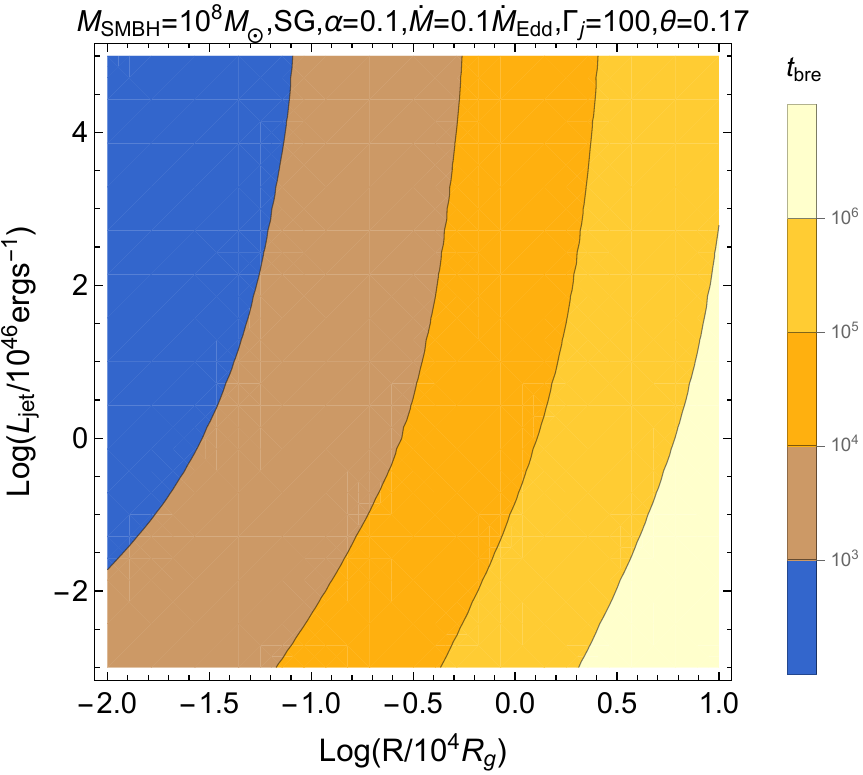}
		\quad
		\includegraphics[width=0.31\textwidth]{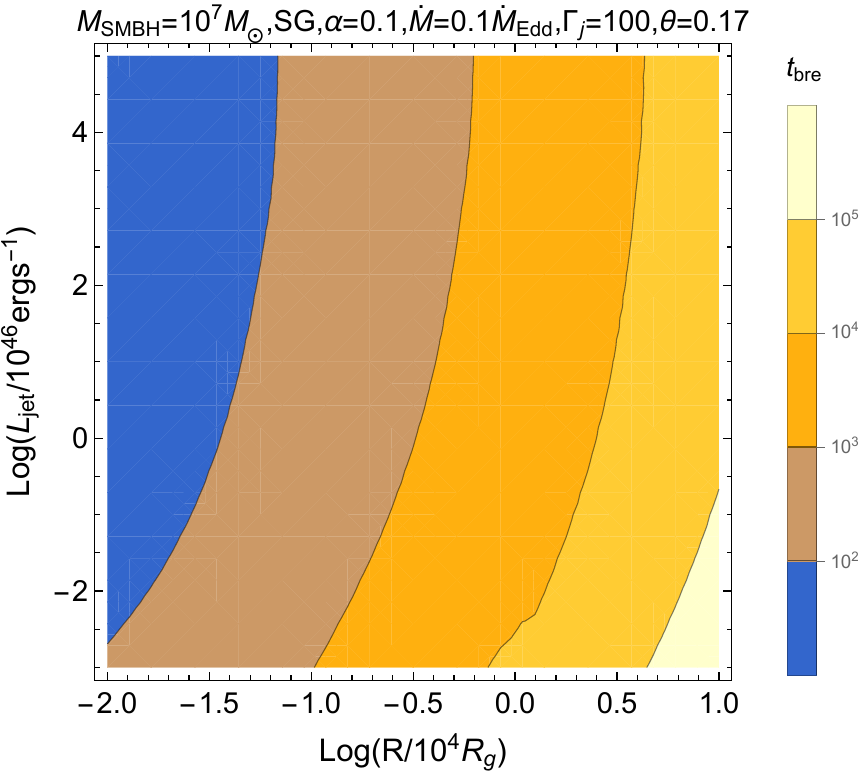}
		\quad
		\includegraphics[width=0.315\textwidth]{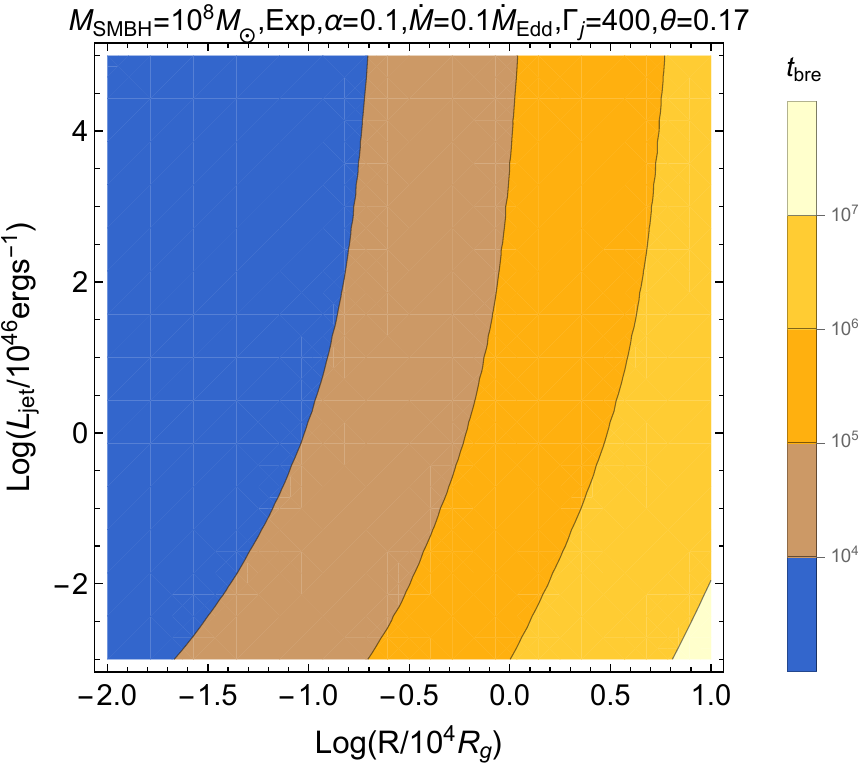}
	\end{center}
	\caption{Time for shock breakout of the embedded-jet, 
		assuming its launch at the midplane of AGN disk, 
		in units of $\text{s}$. 
		The parameters are identical to those in Figure 
		\ref{Fig:bre}. In the middle panel, the SMBH mass is set
	    to $M=10^{7}\text{M}_{\odot}$.}	
	\label{Fig:tbre}
\end{figure*}

We define a critical height $z_{\text{bre}}$ as the jet shock breakout height, where 
$\tau_{\text{d}}(z_{\text{bre}})=1/\beta_{\text{h}}$. Below $z_{\text{bre}}$, photons are 
predominantly trapped within the jet-cocoon system because the photon diffusion 
is slower than the shock propagation, where we simplistically neglect photon leakage 
from the laterally expanding cocoon. Above $z_{\text{bre}}$, photons can efficiently escape 
from the shocked jet-head, producing thermal emission and leading to the jet 
breakout. The width of the shock breakout shell can be estimated as \citep{ChenK24}
\begin{equation}
d_{\text{bre}}=\frac{1}{\kappa\rho(z_{\text{bre}})\beta_{\text{h}}}.
\end{equation}
As shown in Figure \ref{Fig:bre} and Table \ref{Table1}, 
the breakout shell displays an extremely thin structure, 
indicating that jet breakout occurs close to the disk surface, 
attributed to the high jet-head velocity and the substantial 
environmental optical depth. The jet-head velocity varies 
significantly as a function of the jet power, ranging from Newtonian 
to mild-relativistic regimes, which contributes to distinct properties 
in its shock breakout emission (see below). Given that the jet
experiences reacceleration in the isothermal disk, the jet-head velocity 
at breakout exceeds that in a uniform disk.
Correspondingly, the shock breakout time $t_{\text{bre}}$ of the jet can 
be defined as the time required for its head to reach $z_{\text{bre}}$. As shown 
in Figure \ref{Fig:tbre}, a more rapid breakout is observed in the inner region 
of the AGN disk surrounding less massive SMBH, mainly attributed to the thinner 
vertically structure of the disk. Meanwhile, a more powerful jet breaks out 
relatively earlier because it is less prone to deceleration. Additionally,
it is evident that a thicker isothermal or polytropic disk results in a later
breakout.

Depending on the jet-launching mechanism of the central engine, a powerful jet 
can remain active for a duration of $t_{\text{j}}$ before being shut off. 
During propagation through the AGN disk, the jet undergoes deceleration at its 
head while maintaining ultra-relativistic velocities along its body. After 
the central engine stops, the last launched tail continues to pursue the jet 
head until it eventually catches up. If the convergence occurs prior to the jet 
shock breakout, the jet becomes choked and all of its matter and kinetic energy 
are transferred into the cocoon \citep[e.g.][]{Irwin19, Eisenberg22, Pais23}. 
Therefore, the jet choking time is given by
\begin{equation}
t_{\text{ch}}=t_{\text{j}}+\int_{0}^{t_{\text{ch}}}\beta_{\text{h}}\text{d}t.
\end{equation}
If the jet choking time is shorter than the jet breakout time $t_{\text{bre}}$, 
the jet fails to emerge from the AGN disk, leaving only a cocoon; conversely, 
the jet successfully breaks out. A critical situation occurs when 
$t_{\text{ch}}=t_{\text{bre}}$, leading to the derivation of the minimum jet 
duration time as $t_{\text{j,cri}}=t_{\text{bre}}-z_{\text{h}}(t_{\text{bre}})/c$.

\begin{figure*}
	\begin{center}
		\includegraphics[width=0.31\textwidth]{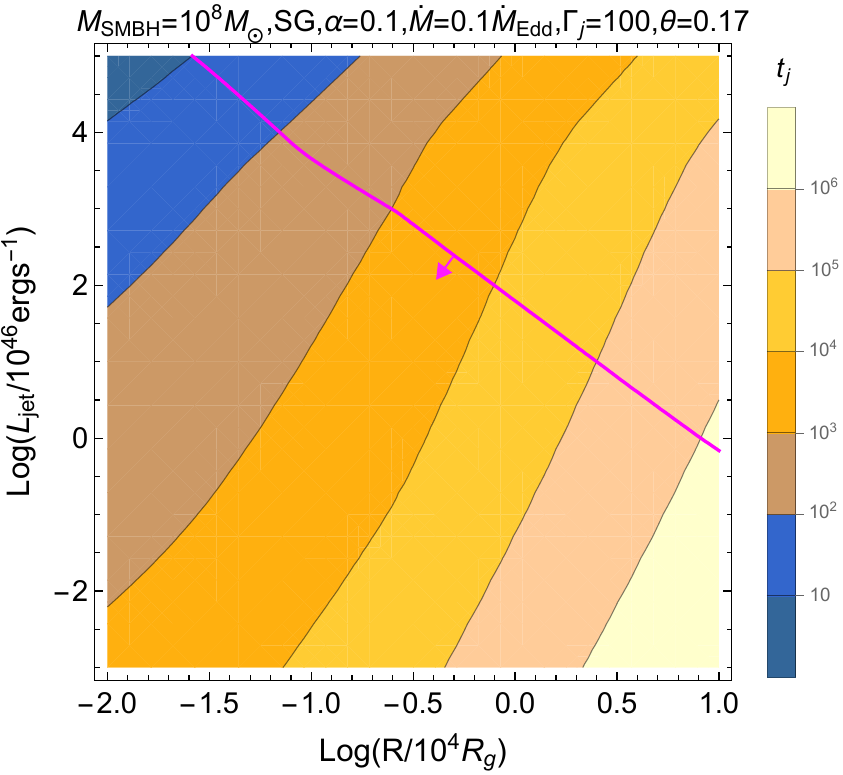}
		\quad
		\includegraphics[width=0.31\textwidth]{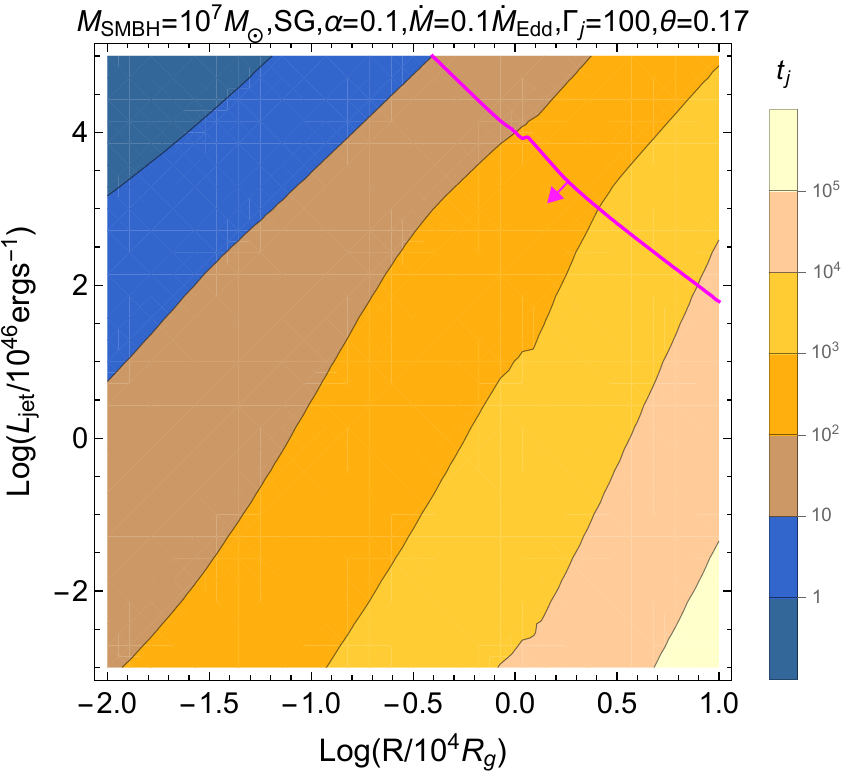}
		\quad
		\includegraphics[width=0.315\textwidth]{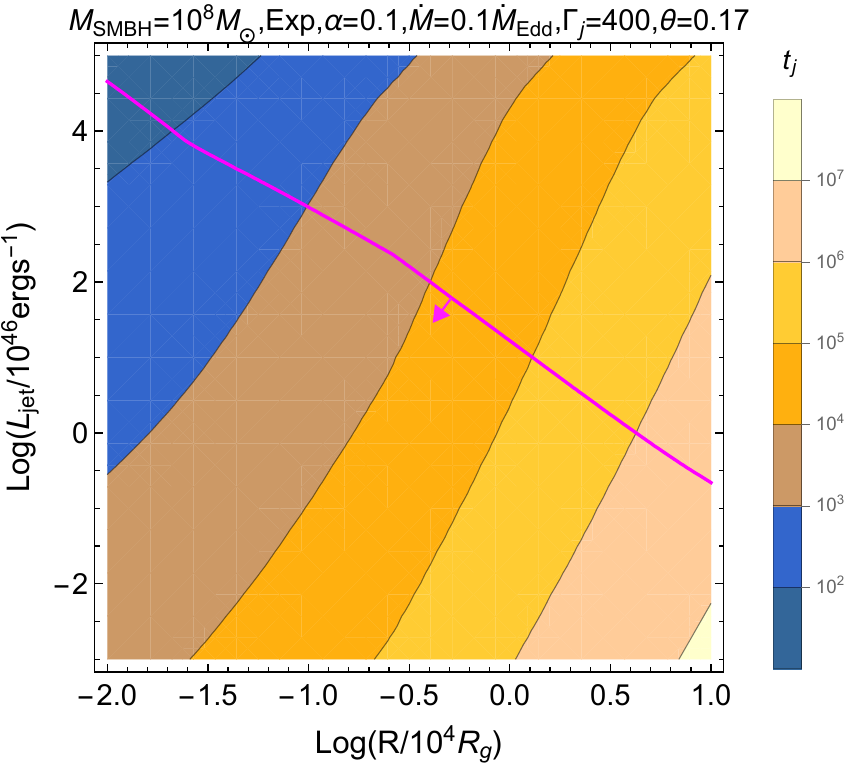}
	\end{center}
	\caption{Minimum active duration of the center engine for jet breakout, 
		in units of $\text{s}$. The parameters
		are identical to those in Figure \ref{Fig:tbre}. For a jet with specific power at a
		given AGN disk location, the left side of the contour represents the choked region. 
		The magenta line and the region to its left indicate that the total energy of the 
		jet is $\leqslant10^{52}\text{\,erg}$.}	
	\label{Fig:tj}
\end{figure*}

The value of $t_{\text{j,cri}}$ is shown in Figure \ref{Fig:tj}. 
Similar to the properties of $t_{\text{bre}}$, we find that 
jets with higher power launched at the inner region of the AGN disk around 
less massive SMBH are more likely to break out. The extended isothermal or
polytropic disk significantly impedes jet propagation, leading to a more 
delayed breakout and a higher $t_{\text{j,cri}}$. Given the substantial variations
in the power, duration, and total energy of jets embedded in AGN disks, 
such as for GRB jets where the duration ranges from $10^{-2}\text{\,s}$ to
$10^{3}\text{\,s}$ \citep[e.g.][]{Qin13} and the total energy primarily spans
$10^{48}\text{\,erg}$ to $10^{53}\text{\,erg}$ \citep[e.g.][]{Liang08, Cenko10},
while for BBH-merger jets or isolated-accreting-BH jets, the power ranges 
from $10^{43}\text{\,erg\,s}^{-1}$ to $10^{50}\text{\,erg\,s}^{-1}$ 
\citep[e.g.][]{Wang21b, Tagawa23a, Tagawa23b, Tagawa24, ChenK24}, with
the duration varying on scale of the accretion timescale over a wide range 
from $10^{4}\text{\,s}$ to $10^{7}\text{\,s}$ \citep[e.g.][]{Tagawa22, ChenK24, Kim24},
these jets may either successfully break out of the AGN disks or become choked. 

\section{Thermal Emission}
\label{section3}
No matter whether the jet is choked or not, the entire jet-cocoon system 
remains opaque when embedded within the AGN disk. The shocked regions, 
specifically the jet head and cocoon, are radiation-mediated where photons 
and gas are tightly coupled. Consequently, the initial emission from the 
breakout system appears as thermal radiation. In what follows, we investigate 
the thermal emission from the jet head, disk cocoon, and jet cocoon when the 
jet successfully breaks out of the AGN disk, or the thermal emission from the 
disk and jet cocoon in the case of a choked jet (see Figure \ref{Fig:sch}).

\subsection{Jet-head Shock Breakout Emission}
The emission properties of a radiation-mediated shock, including luminosity, 
duration, and temperature, are significantly influenced by both its velocity 
and the ambient gas density \citep[e.g.][]{Levinson20}. 
As shown in Figure \ref{Fig:bre}, the velocity of the jet head at shock 
breakout varies significantly among jets with different power levels 
originating from various AGN disk locations with distinct gas densities. 
Consequently, the jet-head shock breakout emission correspondingly exhibits 
diverse properties. Following \cite{Tagawa23b} and \cite{ChenK24}, we 
investigate this emission component.

When the jet head is Newtonian, photons behind the radiation-mediated shock
can escape from the breakout shell and carry away its internal energy within 
the diffusion timescale. According to the shock jump condition \citep{Frank02}, 
internal energy constitutes the major portion of the total energy, so it
is given by
\begin{equation}
	E_{\text{h,bre}}\simeq\rho\Sigma_{\text{h}}d_{\text{bre}}\beta_{\text{h}}^2c^2,
\end{equation}
where all parameter values are evaluated at $z_{\text{bre}}$. Considering the 
diffusion timescale as the duration of the emission, which can be expressed 
as
\begin{equation}
	t_{\text{diff,h}} = 
	\frac{1}{\kappa\rho\beta_{\text{h}}^2c},
\end{equation}
the luminosity of the jet-head shock breakout emission is estimated as
\begin{equation}
L_{\text{h,bre}} \simeq\frac{E_{\text{h,bre}}}{t_{\text{diff,h}}} 
=\Sigma_{\text{h}} \rho \beta_{\text{h}}^3 c^3.
\end{equation}

The typical radiation temperature of the breakout shell is determined by
its degree of thermalization \citep[e.g.][]{Nakar10}. The thermal 
equilibrium temperature behind the radiation-dominant shock is
\begin{equation}\label{Tbbh}
T_{\text{BB,hbre}} = 
\left(\frac{18}{7a}\rho\beta_{\text{h}}^2c^2\right)^{\frac{1}{4}},
\end{equation}
where $a$ is the radiation constant. For a sufficiently high-velocity 
shock, the photons generated through the free-free process are 
inadequate to share the internal energy effectively. 
In this scenario, Compton scattering provides additional photon-electron 
coupling, allowing the breakout shell to approach equilibrium at a 
temperature larger than $T_{\text{BB,hbre}}$ \citep{Nakar10, Katz10}.
The Comptonization-modified radiation temperature within the range 
$0.03<\beta_{\text{h}}<0.4$ can be calculated using the fitting formulae 
presented as \citep{Sapir13}
\begin{equation}\label{eqcom}
\log_{10}\left(\frac{T_{\text{Com,hbre}}}{\text{eV}}\right)=0.975
+1.735\left(\frac{\beta_{\text{h}}}{0.1}\right)^{\frac{1}{2}}
+\left[0.26-0.08\left(\frac{\beta_{\text{h}}}{0.1}\right)^{\frac{1}{2}}\right]
\log_{10}\left(\frac{n}{10^{15}\text{\,cm}^{-3}}\right),
\end{equation}
where $n=\rho/m_{\text{p}}$ is the number density of the AGN disk gas. 
For $\beta_{\text{h}}<0.03$, the
breakout shell reaches thermal equilibrium, and the radiation 
temperature is given by Equation (\ref{Tbbh}).

The thermal emission characteristics of a relativistic breakout shell 
exhibit substantial differences compared to those in the aforementioned 
Newtonian shocked shell. For $\beta_{\text{h}}>0.5$, even when accounting 
for Compton scattering, the shock temperature exceeds $50\, \text{keV}$,
leading to the copious production of positron-electron pairs 
\citep[e.g.][]{Weaver76, Svensson82}. These pairs are the primary sources 
that generate photons, facilitating the shock downstream to reach 
equilibrium, and adjust the rest-frame temperature behind a 
radiation-mediated shock to be nearly constant at $\sim200\, \text{keV}$ 
\citep[e.g.][]{Budnik10, Katz10}. Meanwhile, pairs production 
substantially increases the leptons number density within the breakout 
shell, rendering it initially opaque even though the upstream optical depth 
$\tau_{\text{d}}(z_{\text{bre}})\leqslant1$ \citep[e.g.][]{Budnik10, Nakar12}.
Moreover, after breaking out of the AGN disk, the relativistic shell 
undergo continuous acceleration driven by its internal energy until 
either all energy is converted into bulk kinetic energy or the shell 
becomes transparent \citep[e.g.][]{Tan01, Pan06}. We follow \cite{Nakar12}
to model the emission.

The initial internal energy of the breakout shell is 
\begin{equation}
	E_0=\gamma_{\text{h}}^2M_{\text{h}}c^2,
\end{equation}
where $\gamma_{\text{h}}=\Gamma_{\text{h}}/\sqrt{2}$ is the initial Lorentz factor
of the shocked gas, and 
$M_{\text{h}}=\Sigma_{\text{h}}d_{\text{bre}}\rho\sim\Sigma_{\text{h}}/\kappa$ is the 
mass of the shell. Considering shell acceleration, pair-production, and 
adiabatic cooling, photons can diffuse out of the shell at $t_{\text{h,th}}$,
by which time the shell has attained $\gamma_{\text{h,th}}$, with the specific
expressions detailed in Appendix \ref{Appendix-Rel-shell}. The emission 
energy is given by
\begin{equation}
	E_{\text{h,bre}}^{\text{R}}=E_0\frac{\gamma_{\text{h,th}}}{\gamma_{\text{h}}}
	\frac{T_{\text{th}}^{'}}{T_{\text{h}}^{'}},
\end{equation}
where $T_{\text{h}}^{'}=200\, \text{keV}$ and $T_{\text{th}}^{'}=50\, \text{keV}$ are 
the rest-frame temperatures at shell breakout and when the shell becomes
optically thin \citep{Nakar12}, respectively. The observed duration of 
the emission can be estimated as
\begin{equation}
	t_{\text{h,bre}}^{\text{obs}}\approx\frac{t_{\text{h,th}}}{2\gamma_{\text{h,th}}^2},
\end{equation}
and the resulting luminosity of the jet-head shock breakout emission is
\begin{equation}
	L_{\text{h,bre}}^{\text{R}} 
	\simeq\frac{E_{\text{h,bre}}^{\text{R}}}{t_{\text{h,bre}}^{\text{obs}}} 
	=\frac{\gamma_{\text{h,th}}^3}{2\gamma_{\text{h}}}\frac{E_0}{t_{\text{h,th}}}.
\end{equation}
Considering the beaming effect, the isotropic equivalent luminosity would 
correspond to an amplification of $L_{\text{h,bre}}^{\text{R}}$ by a factor of 
$\sim2\gamma_{\text{h,th}}^2$. For the relativistic shell, the observed
temperature of the breakout emission is
\begin{equation}
	T_{\text{h,bre}}^{\text{obs}}=T_{\text{th}}^{'} \gamma_{\text{h,th}}.
\end{equation}
For simplicity, we extrapolate the velocity limit and apply the relativistic 
shell model for cases where $\beta_{\text{h}} > 0.4$.

\subsection{Disk-cocoon Emission}
\label{section-disk-cocoon}
After the jet-head shock breakout, no further energy is injected into the 
cocoon. Subsequently, the radiation-dominated disk cocoon expands under 
its inherent pressure. Although overly simplistic, we independently 
investigate the evolution of cocoon to obtain a preliminary understanding 
of its thermal emission properties. The initial parameters of the cocoon 
are determined by the jet-cocoon dynamics at the jet breakout. We assume 
half of the cocoon energy is allocated to its disk-component.

Given the longer evolutionary timescale of the disk cocoon compared to the 
jet-head, we investigate the temporal evolution of luminosity for disk-cocoon
emission, assuming that this emission arises subsequent to the jet-head 
shock breakout. Similar to the jet head, the disk cocoon emits 
radiation during its shock breakout, with the luminosity and timescale 
of the emission given by
\begin{equation}
	L_{\text{c,bre}} =\pi r_{\text{c}}^2 \rho \beta_{\text{c}}^3 c^3, \label{Lcbre}
\end{equation}
and
\begin{equation}
t_{\text{c,bre}} = 
\frac{1}{\kappa \rho \beta_{\text{c}}^2 c}\label{tcbre}.
\end{equation}
The luminosity at $t \leq t_{\text{c,bre}}$ is approximated by the 
leakage of photons from a static slab, characterized by a diffusion 
time $t_{\text{c,bre}}$ \citep{Nakar10}, with the rise in luminosity 
following $L_{\rm{c}}\propto \exp[1-t_{\text{c,bre}}/t] $; 
while for $t_{\text{c,bre}} < t \leq t_{\text{c,pla}}$, the breakout 
shell undergoes adiabatic expansion primarily perpendicular 
to the disk surface to grow its width in the planar phase, where
$t_{\text{c,pla}}$ denotes the transition time between the planar 
and spherical geometries of the expanding cocoon. The 
planar geometry dictates the luminosity evolving as
$L_{\rm{c}}\propto (t/t_{\text{c,bre}})^{-\frac{4}{3}} $ \citep{Nakar10}, 
with the simplification neglecting the light travel effect. 
For $t > t_{\text{c,pla}}$, the expansion geometry evolves into a 
nearly spherical configuration with an initial radius 
$\sim r_{\text{c}}$. The luminosity is modeled as the expansion of 
an opaque fireball without external power input,
starting from $t_{\text{c,pla}}$, in accordance with 
Equation (3) in \cite{Chatzopoulos12}. At the beginning 
of spherical phase, the internal energy of the cocoon material 
escaping the AGN disk is approximate 
$E_{\text{c,th0}}\sim E_{\text{c}} r_{\text{c}}/2H_{\text{d}}$, 
and the initial diffusion timescale is 
$t_{\text{c,0}}=\kappa (r_{\text{c}}/H_{\text{d}}) M_{\text{c}}/b c r_{\text{c}}$, 
where $M_{\text{c}}=\rho V_{\text{c}}$ is the mass of the disk-cocoon, and 
$b=13.8$ \citep{Arnett82}. The resulting effective diffusion
timescale is thus \citep{Chatzopoulos12}
\begin{equation}\label{tcdiff}
t_{\text{c,diff}}=\sqrt{t_{\text{c,0}}t_{\text{s,0}}}
=\sqrt{\frac{2}{\pi b}}\frac{\kappa M_{\text{c}}}{c H_{\text{d}}},
\end{equation}
where $t_{\text{s,0}}=r_{\text{c}}/\beta_{\text{c}}c$. 
And the initial luminosity in the spherical phase is
\citep{Chatzopoulos12}
\begin{equation}\label{Lcsph}
L_{\text{c,sph}}=\frac{E_{\text{c,th0}}}{t_{\text{c,0}}}=
\frac{b c r_{\text{c}} E_{\text{c}}}{2 \kappa M_{\text{c}}}.
\end{equation}
Since the structure of disk-cocoon is cylindrical rather than 
spherical, and its evolution resembles a bubble rising and 
expanding on the surface of the AGN disk, the luminosity 
evolution is unlikely to fully adhere to an isotropic spherical 
expansion \citep[e.g.][]{Nakar14}. Therefore, the actual bolometric
light curve should be examined via numerical simulation methods. 
However, for simplicity, we directly connect the planar and 
spherical phase at
\begin{equation}
t_{\text{c,pla}}=t_{\text{c,bre}}\left(\frac{L_{\text{c,bre}}}
{L_{\text{c,sph}}}\right)^{\frac{3}{4}}.
\end{equation}
Consequently, the temporal evolution of luminosity for disk-cocoon
emission is described by
\begin{equation}\label{Lcnacc}
L_{\text{c}}(t) = 
\begin{cases}
L_{\text{c,bre}}\exp\left[1-\frac{t_{\text{c,bre}}}{t}\right], & t \leq t_{\text{c,bre}} \\[1.5ex]
L_{\text{c,bre}}\left(\frac{t}{t_{\text{c,bre}}}\right)^{-\frac{4}{3}}, & t_{\text{c,bre}} < t \leq t_{\text{c,pla}} \\[1.5ex]
L_{\text{c,sph}}\exp\left[-\frac{(t-t_{\text{c,pla}})(t+t_{\text{c,pla}})}{2t_{\text{c,diff}}^2}\right]. & t > t_{\text{c,pla}}
\end{cases}
\end{equation}

For $\beta_{\text{c}}<0.03$, the disk-cocoon maintains thermal 
equilibrium. The temporal evolution of the emission temperature
can be derived from the observed effective temperature
$T_{\text{BB,c}} \propto (L_{\text{c}}/r_{\text{ph}})^{\frac{1}{4}}$
\citep[e.g.][]{Piro21},
where the photosphere radius $r_{\text{ph}}$ remains nearly 
constant during the breakout and planar phases, whereas
$r_{\text{ph}}\simeq \beta_{\text{c}} t$ during the spherical
phase. Therefore, the emission temperature evolves as 
\begin{equation}\label{TcnaccBB}
T_{\text{BB,c}}(t) \propto
\begin{cases}
\exp\left[1-\frac{t_{\text{c,bre}}}{t}\right]^{\frac{1}{4}}, 
& t \leq t_{\text{c,bre}} \\[1.5ex]
\left(\frac{t}{t_{\text{c,bre}}}\right)^{-\frac{1}{3}}, 
& t_{\text{c,bre}} < t \leq t_{\text{c,pla}} \\[1.5ex]
\exp\left[-\frac{(t-t_{\text{c,pla}})
(t+t_{\text{c,pla}})}{2t_{\text{c,diff}}^2}\right]^{\frac{1}{4}}
\left(\frac{t}{t_{\text{c,pla}}}\right)^{-\frac{1}{2}}, & t > t_{\text{c,pla}}
\end{cases}
\end{equation}
where $T_{\text{BB,c}}(t_{\text{c,bre}}) = T_{\text{BB,cbre}}=
\left(18\rho\beta_{\text{c}}^2c^2/7a\right)^{\frac{1}{4}}$. 

When $\beta_{\text{c}}>0.03$, the emission temperature
is initially out of thermal equilibrium, leading to a more 
complex evolution process.
The degree of thermalization can be reflected by a coefficient
$\eta \equiv n_{\text{BB}}/t_{\text{cp}} \dot{n}_{\text{ph,ff}}$
\citep{Nakar10}, where 
$n_{\text{BB}}\sim aT_{\text{BB}}^4/3kT_{\text{BB}}$ 
is the photon number density required for 
thermal equilibrium, $k$ is the Boltzmann constant, 
$\dot{n}_{\text{ph,ff}} \approx 3.5\times 10^{36}
\text{\,s}^{-1}\text{\,cm}^{-3} \rho^2 T^{-\frac{1}{2}}$ is the 
photon production rate via free-free process, and
$t_{\text{cp}}$ is the time over which photons can couple 
with the gas. For $\eta >1$, the Comptonization-modified
temperature is given by $T_{\text{Com,c}}=T_{\text{BB,c}}
\eta(T_{\text{BB,c}})^2/\xi(T_{\text{Com,c}})^2$ \citep{Nakar10},
where $\xi$ is the Comptonization correction factor. 
Neglecting the evolution of $\xi$, an approximate relation 
$T_{\text{Com,c}}\propto T_{\text{BB}}\eta^2$ is derived. 
During the breakout phase, accounting for photon 
leakage from a static slab \citep{Nakar10}, 
the optical depth of the breakout shell
$\tau \propto t$, the thermal equilibrium temperature
$T_{\text{BB}} \propto L_{\text{c}} \tau$, and 
$t_{\text{cp}} \propto t^2$, resulting in
$\eta \propto \exp \left[1-t_{\text{c,bre}}
/t\right]^{\frac{7}{8}} t^{-\frac{9}{8}}$,
and thereby $T_{\text{Com,c}}\propto 
\exp[1-t_{\text{c,bre}}/t]^{2}(t/t_{\text{c,bre}})^{-2}$. 
Similarly, during the planar phase, the observed temperate
evolves as $T_{\text{Com,c}}\propto(t/t_{\text{c,bre}})^{-\frac{2}{3}}$ 
\citep{Nakar10}. In the early
stage of spherical phase, the increase in $\eta$ leads 
to inefficient photon production. The temperature thus 
evolves following adiabatic cooling, 
$T_{\text{Com,c}} \propto (\beta_{\text{c}}^3 t^3)^{-\frac{1}{3}}$, 
until at $t_{\text{c,th}}$, the expanding disk-cocoon
re-establishes thermal equilibrium with 
$T_{\text{Com,c}}= 
(L_{\text{c}}/4\pi r_{\text{ph}}^2\sigma_{\text{SB}})^{\frac{1}{4}}$, 
where $\sigma_{\text{SB}}$ is the Stefan–Boltzmann constant.
Combining all phases, the emission temperature can be approximately 
described as
\begin{equation}\label{TcnaccCom}
T_{\text{Com,c}}(t) \propto
\begin{cases}
\exp\left[1-\frac{t_{\text{c,bre}}}{t}\right]^{2}\left(\frac{t}{t_{\text{c,bre}}}\right)^{-2}, & t \leq t_{\text{c,bre}} \\[1.5ex]
\left(\frac{t}{t_{\text{c,bre}}}\right)^{-\frac{2}{3}}, & t_{\text{c,bre}} < t \leq t_{\text{c,pla}} \\[1.5ex]
\left(\frac{t}{t_{\text{c,pla}}}\right)^{-1}, & t_{\text{c,pla}} < t \leq t_{\text{c,th}} \\[1.5ex]
\exp\left[-\frac{(t-t_{\text{c,th}})(t+t_{\text{c,th}})}{2t_{\text{c,diff}}^2}\right]^{\frac{1}{4}}\left(\frac{t}{t_{\text{c,th}}}\right)^{-\frac{1}{2}}, &  t > t_{\text{c,th}}
\end{cases}
\end{equation}
where $T_{\text{Com,c}}(t_{\text{c,bre}}) = T_{\text{Com,cbre}}$, 
which is calculated via substituting $\beta_{\text{c}}$ in 
Equation (\ref{eqcom}) to replace $\beta_{\text{h}}$.
Furthermore, during the planar phase, the photon-gas thermal 
coupling increases over time as $\eta \propto t^{-\frac{1}{6}}$ 
\citep{Nakar10}. Consequently, the breakout shell may 
attain thermal equilibrium when $\eta=1$ at 
$t_{\text{c,thp}}=\eta(t_{\text{c,bre}})^6 t_{\text{c,bre}}$
in the planar phase. If $t_{\text{c,thp}}<t_{\text{c,pla}}$, 
the emission temperature evolves as
\begin{equation}\label{Tcomp}
T_{\text{Com,c}}(t) \propto
\begin{cases}
\exp\left[1-\frac{t_{\text{c,bre}}}{t}\right]^{2}\left(\frac{t}{t_{\text{c,bre}}}\right)^{-2}, & t \leq t_{\text{c,bre}} \\[1.5ex]
\left(\frac{t}{t_{\text{c,bre}}}\right)^{-\frac{2}{3}}, & t_{\text{c,bre}} < t \leq t_{\text{c,thp}} \\[1.5ex]
\left(\frac{t}{t_{\text{c,thp}}}\right)^{-\frac{1}{3}}, & t_{\text{c,thp}} < t \leq t_{\text{c,pla}} \\[1.5ex]
\exp\left[-\frac{(t-t_{\text{c,pla}})(t+t_{\text{c,pla}})}{2t_{\text{c,diff}}^2}\right]^{\frac{1}{4}}\left(\frac{t}{t_{\text{c,pla}}}\right)^{-\frac{1}{2}}. & t > t_{\text{c,pla}}
\end{cases}
\end{equation}

Next, we investigate the luminosity and temperature evolution 
of the disk-cocoon emission in the stratified disk.
As shown in Figure \ref{Fig:dyn}, the disk cocoon undergoes 
continuous deceleration because, according to Equation (\ref{eq:bc}), 
$\beta_{\text{c}} \propto \overline \rho(z_{\text{h}})^{-\frac{1}{2}}$,
where the vertical averaged density of both isothermal and 
polytropic disk is determined by the heavy material
within $H_{\text{d}}$ and decreases slowly. Although employing
the averaged density provides a more precise description on jet 
collimation and the general dynamical evolution of the 
jet-cocoon system, it significantly overestimates the 
local density at heights exceeding $H_{\text{d}}$, 
which determines the instantaneous velocity of the cocoon. 
As a result, the sharp drop in density above $H_{\text{d}}$ in an
isothermal or polytropic disk results in the acceleration
of the uppermost portion of the disk cocoon material. 
This acceleration induces additional evolution of luminosity 
and temperature in the early stages following breakout 
\citep[e.g.][]{Nakar10, Piro21}. Detailed properties of this 
disk-cocoon emission are provided in Appendix 
\ref{Appendix-Disk-Cocoon}.

\subsection{Jet-cocoon Emission}
\label{section-jet-cocoon}
We follow \cite{Nakar17} to investigate the jet-cocoon emission.
During the jet propagation within the AGN disk, the less-dense 
jet-cocoon would undergo mixing with the disk-cocoon via 
instabilities \citep[e.g.][]{Morsony07, Gottlieb20, Gottlieb21}, 
despite the radiation-dominated pressure (energy density) being 
equal \citep{Bromberg11}. Depending on the interaction duration,
shocked jet materials at different heights hold varying 
mixing level, and thereby jet-cocoon becomes stratified, with
material near the base of jet showing greater mixing and higher
density, while material closer to the jet head experiences 
less mixing and lower density \citep{Eisenberg22}. Moreover,
mixing results in an approximately uniform distribution of jet-cocoon
energy on a logarithmic scale of energy per baryon, primarily 
in the form of internal energy \citep{Gottlieb20, Gottlieb21}.
Consequently, the fraction of jet-cocoon energy deposited in 
material with a specific terminal proper velocity 
$\Gamma_{\text{cj}} \beta_{\text{cj}}$, after internal energy primarily 
converted into kinetic energy, can be assumed to a constant
$f_{\Gamma \beta}$, denoted as $f_{\Gamma \beta} \sim 0.1$ 
\citep{Nakar17}.

For the Newtonian jet-cocoon material with a terminal velocity
$\beta_{\text{cj}} \lesssim 0.7$, its mass is estimated as
\begin{equation}
m_{\text{cj},\beta} \simeq \frac{2 f_{\Gamma \beta} E_{\text{cj}}}{\beta_{\text{cj}}^2 c^2},
\end{equation}
where $E_{\text{cj}}\simeq E_{\text{c}}/2$ is the total energy of the jet-cocoon, 
and its initial volume can be approximated by 
\begin{equation}
V_{\text{cj},\beta} \simeq f_{\Gamma \beta} \pi r_{\text{c}}^2 z_{\text{bre}}.
\end{equation}
After adiabatic expansion, photons can diffuse out of the jet-cocoon
shell, referred to as the luminosity shell \citep{Nakar10}, to generate 
emission when $\kappa m_{\text{cj},\beta}/4\pi r_{\text{cj}}^2=1/\beta_{\text{cj}}$ 
is satisfied, thus the diffusion radius is given by
\begin{equation}
r_{\text{cj}}=\left(\frac{\kappa f_{\Gamma \beta} E_{\text{cj}}}
{2\pi\beta_{\text{cj}}c^2}\right)^{\frac{1}{2}},
\end{equation}
and the corresponding diffusion time is 
\begin{equation}
t_{\text{cj}} \simeq 
\left(\frac{\kappa f_{\Gamma \beta} E_{\text{cj}} }
{2\pi \beta_{\text{cj}}^3 c^4}\right)^{\frac{1}{2}}.
\end{equation}
The resulting emission luminosity is estimated as 
\begin{equation}\label{Lcjb}
L_{\text{cj}}
=\frac{f_{\Gamma \beta} E_{\text{cj}}}
{t_{\text{cj}}}\frac{V_{\text{cj},\beta}^{\frac{1}{3}}}{r_{\text{cj}}}
=\frac{2 \pi \beta_{\text{cj}}^2 c^3 V_{\text{cj},\beta}^{\frac{1}{3}}}{\kappa}.
\end{equation}
We set $\beta_{\text{cj,s}} = \text{min}[0.7,\beta_{\text{cr}}]$ 
as the velocity of the first shell that produces diffusing 
thermal emission, where 
$\beta_{\text{cr}} = \kappa f_{\Gamma \beta} E_{\text{cj}}/2\pi r_{\text{c}}^2 c^2$ 
represents the critical velocity at which $r_{\text{cj}}=r_{\text{c}}$, 
signifying that
photons escape from the jet-cocoon just after the jet breakout
\citep{ChenK24}. Based on the energy distribution 
$\text{d}E/\text{d}\log\beta_{\text{cj}}=f_{\Gamma \beta} E_{\text{cj}}$,
the mass distribution can be described as
$m(>v)\propto v^{-(s+1)}$, where $s\simeq 1$ \citep{Nakar17}. Therefore,
the properties of luminosity shell evolve over time as 
$m_{\text{cj}} \propto t^{\frac{2(s+1)}{s+2}}$, 
$\beta_{\text{cj}} \propto t^{-\frac{2}{s+2}}$,
$r_{\text{cj}} \propto t^{\frac{s}{s+2}}$, and  
$\tau_{\text{cj}} \propto t^{\frac{2}{s+2}}$. As the emission luminosity 
$L_{\text{cj}} \propto \beta_{\text{cj}}^2$, it evolves as
\begin{equation}
L_{\text{cj}}(t) = L_{\text{cj,s}}
\begin{cases}
\left(\frac{t}{t_{\text{cj,s}}}\right)^{-\frac{4}{s+2}}, & t_{\text{cj,s}} < t \leq t_{\text{sph,end}} \\[1.5ex]
\left(\frac{t_{\text{sph,end}}}{t_{\text{cj,s}}}\right)^{-\frac{4}{s+2}}\exp\left[-\frac{1}{2}\left(\frac{t^2}{t_{\text{sph,end}}^2}-1\right)\right], & t > t_{\text{sph,end}}
\end{cases}
\end{equation}
where $L_{\text{cj,s}}$ and $t_{\text{cj,s}}$ are calculated at 
$\beta_{\text{cj,s}}$; $t_{\text{sph,end}}$ denotes the time 
at which the innermost fully mixed jet-cocoon, whose
velocity is assumed to be $\beta_{\text{c}}$, becomes radiative, 
i.e., $t_{\text{sph,end}}=
t_{\text{cj,s}}(\beta_{\text{cj,s}}/\beta_{\text{c}})^{\frac{s+2}{s}}$. 
At $t>t_{\text{sph,end}}$, the emission luminosity decreases exponentially
as the shell undergoes simultaneous radiative cooling and adiabatic 
expansion \citep{Piro21}. The temporal evolution of the emission 
temperature is examined in a manner analogous to that of the 
disk-cocoon, which is presented in Appendix \ref{TcjN}.

For relativistic jet-cocoon materials, the properties of emission
are determined by a combination of shell acceleration, spreading, and
thermalization \citep{Nakar17}. In the dilute radiation-dominated 
jet-cocoon material, photons can accelerate baryons to a critical 
Lorentz factor before decoupling, which is given by 
\citep{Nakar05, Nakar17}
\begin{equation}
\gamma_{\text{b}}=\left(\frac{E_{\text{cj}}\sigma_{\text{T}}}
{4\pi z_{\text{bre}}^2\theta_{\text{cj}}^2\theta_{\text{c}}m_{\text{p}}c^2}\right)^{\frac{1}{4}}
=5.5 E_{\text{cj,50}}^{\frac{1}{4}}z_{\text{bre,13}}^{-\frac{1}{2}} 
\theta_{\text{cj}}^{-\frac{1}{2}}\theta_{\text{c}}^{-\frac{1}{4}},
\end{equation}
where the numbers in subscripts represent the power of 10, e.g., 
$E_{\text{cj,50}}=E_{\text{cj}}/10^{50}\text{\,erg}$; and
$\theta_{\text{c}}=r_{\text{c}}/z_{\text{bre}}$, 
$\theta_{\text{cj}}\simeq0.5$ are the initial and final 
opening angle of the expanding relativistic jet-cocoon.
Another critical Lorentz factor that describes
the spreading status of the jet-cocoon material is
given by \citep{Nakar05, Nakar17}
\begin{equation}
\gamma_{\text{s}}=\left(\frac{E_{\text{cj}}\sigma_{\text{T}}}
{4\pi f_{\Gamma \beta} z_{\text{bre}}^2\theta_{\text{cj}}^2 m_{\text{p}}c^2}\right)^{\frac{1}{5}}
=4.9 E_{\text{cj,50}}^{\frac{1}{5}}z_{\text{bre,13}}^{-\frac{2}{5}} f_{\Gamma \beta,-1}^{-\frac{1}{5}}
\theta_{\text{cj}}^{-\frac{2}{5}},
\end{equation}
for gas with $\gamma<\gamma_{\text{s}}$, photons decouple from it
when its width is spreading, while for gas with 
$\gamma>\gamma_{\text{s}}$, photon decoupling occurs prior to the 
onset of spreading. Furthermore, by setting the thermal coupling 
coefficient $\eta=1$, where
$T_{\text{BB,cj0}}= (f_{\Gamma \beta}E_{\text{cj}}/a V_{\text{cj},\beta})^{\frac{1}{4}}$,
$\rho_{\text{cj},\gamma}=f_{\Gamma \beta}E_{\text{cj}}/\gamma c^2 V_{\text{cj},\beta}$,
$t_{\text{cp}}\simeq f_{\Gamma \beta}z_{\text{bre}}/c$, the resulting 
coefficient is
\begin{equation}
\eta(\gamma)=1.7\times 10^4 E_{\text{cj,50}}^{-\frac{9}{8}} z_{\text{bre,13}}^{\frac{19}{8}}
f_{\Gamma \beta,-1}^{-1} \theta_{\text{c}}^{\frac{9}{4}} \gamma_{1}^{2},
\end{equation}
and the thermal critical Lorentz factor can 
be expressed as
\begin{equation}
\gamma_{\text{BB}}=0.007 E_{\text{cj,50}}^{\frac{9}{16}}z_{\text{bre,13}}^{-\frac{19}{16}} f_{\Gamma \beta,-1}^{\frac{1}{2}}
\theta_{\text{c}}^{-\frac{9}{8}},
\end{equation}
gas with $\gamma<\gamma_{\text{BB}}$ can reach thermal equilibrium
before generating emission.

In the case of $\gamma_{\text{b}}\lesssim1.4$, photons decouple
from the rarefied gas before accelerating it to relativistic
velocity, leading to most of the internal energy being carried 
away by radiation. Therefore, the duration of the emission is 
estimated as \citep{Nakar17}
\begin{equation}
t_{\text{cj}}\simeq\frac{f_{\Gamma \beta}z_{\text{bre}}}{c},
\end{equation}
and its isotropic equivalent luminosity is
\begin{equation}
L_{\text{cj}}\simeq\frac{E_{\text{cj}}c}{z_{\text{bre}}\theta_{\text{cj}}^2},
\end{equation}
where it is assumed that the photons within jet-cocoon materials 
close to the jet head can escape, while those in the Newtonian 
jet-cocoon materials remain trapped.
Meanwhile, $\gamma_{\text{BB}}\ll 1$ indicates that radiation is 
far from thermal equilibrium. Regulated by electron-positron 
pairs, the observed temperature is approximately 
$T_{\text{cj}}\simeq200\, \text{keV}$.

When $\gamma_{\text{b}}>1.4$, photons can effectively accelerate 
the gas to a terminal Lorentz factor provided that $\gamma<\gamma_{\text{b}}$.
After acceleration, photon-gas decoupling occurs at the photosphere
radius
\begin{equation}
R_{\gamma,\text{ph}}=\left(\frac{\kappa f_{\Gamma \beta} E_{\text{cj}}}
{2\pi \theta_{\text{cj}}^2 c^2 \gamma }\right)^{\frac{1}{2}},
\end{equation}
and thereby the observed duration is
\begin{equation}\label{tcjgamma}
t_{\text{cj},\gamma}=\frac{R_{\gamma,\text{ph}}}{2 c \gamma^2}
=7.5\text{\,s}\, E_{\text{cj,50}}^{\frac{1}{2}}f_{\Gamma \beta,-1}^{-\frac{1}{2}}
\theta_{\text{cj}}^{-1}\gamma_{1}^{-\frac{5}{2}}.
\end{equation}
Given that $t_{\text{cj},\gamma}$ for gas with large $\gamma$ is 
relatively short, we focus on the gas with
$\gamma\leqslant10$ when $\gamma_{\text{b}}>10$. Additionally, 
since $\gamma_{\text{s}}$ is generally slightly smaller than 
$\gamma_{\text{b}}$ across extensive parameter regions, we 
restrict our consideration of the maximum gas Lorentz factor 
to
\begin{equation}
\gamma_{\text{max}}=\text{min}\left[\gamma_{\text{s}},10\right],
\end{equation}
and set the minimum gas Lorentz factor as 
$\gamma_{\text{min}}=3$.
When reaching the photosphere radius, the comoving volume 
of the luminosity shell with Lorentz factor  $\gamma$ 
expands to $V_{\text{ph},\gamma}^{'}\sim 2\pi \theta_{\text{cj}}^2 
R_{\gamma,\text{ph}}^3/\gamma $.
Due to adiabatic cooling, the isotropic 
equivalent luminosity is estimated as
\begin{equation}
L_{\text{cj}}=\frac{f_{\Gamma \beta} E_{\text{cj}}}{t_{\text{cj},\gamma} \theta_{\text{cj}}^2}
\left(\frac{f_{\Gamma \beta} V_{\text{c}}}{V_{\text{ph},\gamma}^{'}}\right)^{\frac{1}{3}}\gamma,
\end{equation}
which evolves over time as $L_{\text{cj}}\propto t^{-\frac{26}{15}}$, 
because of the Lorentz factor of the luminosity shell
following $\gamma \propto t^{-\frac{2}{5}}$. 

The observed temperature is determined by $\gamma_{\text{BB}}$. 
For $\gamma_{\text{BB}}>\gamma_{\text{max}}$, the relativistic
jet-cocoon materials are completely in thermal equilibrium, 
and the temperature is given by
\begin{equation}
T_{\text{cj,obs}}=T_{\text{cj0}}
\left(\frac{f_{\Gamma \beta} V_{\text{c}}}{V_{\text{ph},\gamma}^{'}}\right)^{\frac{1}{3}}\gamma,
\end{equation}
where $T_{\text{cj0}}=T_{\text{BB,cj0}}$. 
As $T_{\text{cj,obs}}\propto \gamma^{\frac{11}{6}}$, the observed 
temperature evolves over time as 
$T_{\text{cj,obs}}\propto t^{-\frac{11}{15}}$. 
On the contrary, for $\gamma_{\text{BB}}<1.4$, throughout the 
evolution of relativistic jet-cocoon materials, the emission 
remains out of thermal equilibrium, and the temperature 
properties can be analyzed similarly to Appendix \ref{TcjN}.
Based on $T_{\text{BB,cj0}}$, $\eta(\gamma_{\text{max}})$, and
pair production, the initial temperature can be determined 
as $T_{\text{cj0}}=\max[T_{\text{Com,cj}},200\,\text{keV}]$. If
$T_{\text{cj0}}=T_{\text{Com,cj}}$, as $\eta\propto\gamma^2$,
the observed temperature evolves as 
$T_{\text{cj,obs}}\propto\gamma^{\frac{35}{6}}
\propto t^{-\frac{7}{3}}$. Otherwise, if 
$T_{\text{cj0}}=200\,\text{keV}$, the initial temperature 
remains constant. When 
$4f_{\Gamma \beta} V_{\text{c}}<V_{\text{ph},\gamma}^{'}$,
photons escape from the photosphere, and 
$T_{\text{cj,obs}}\propto t^{-\frac{11}{15}}$; conversely, 
when $4f_{\Gamma \beta} V_{\text{c}}>V_{\text{ph},\gamma}^{'}$,
the temperature evolution is initially governed by pairs, 
such that $T_{\text{cj,obs}}\propto\gamma\propto t^{-\frac{2}{5}}$,
before being determined by photosphere. Once the initial
temperature decreases below $50\,\text{keV}$ and the pairs
disappear, the evolution reverts to Comptonization.
Additionally, for 
$\gamma_{\text{min}}<\gamma_{\text{BB}}<\gamma_{\text{max}}$,
the evolution of the observed temperature transitions 
from a state out of thermal equilibrium to one in 
thermal equilibrium, with the transition occurring at
$t_{\text{cj},\gamma}(\gamma_{\text{BB}})$. 

\subsection{Choked Jet}

After choking, the jet ceases to exist, and the jet-cocoon system 
exclusively transitions to cocoon expansion driven by its internal pressure.
The initial height and width of the cocoon at $t_{\text{ch}}$ can be 
determined from the jet-cocoon dynamics, i.e., $z_{\text{h}}(t_{\text{ch}})$
and $r_{\text{c}}(t_{\text{ch}})$. The internal energy of the cocoon is given by
$E_{\text{ch}}=L_{\text{j}}t_{\text{j}}$, where the total jet kinetic energy has 
been injected into the cocoon. Unlike the anisotropic jet-cocoon 
evolution where $\beta_{\text{h}}\gg\beta_{\text{c}}$, under the nearly uniform 
pressure approximation \citep[e.g.][]{Bromberg11}, the cocoon exhibits
comparable expansion velocities in both the forward and lateral directions, 
as well as similar expansion scales in these directions \citep{Irwin19}. 
Therefore, the height, width and velocity of the remaining cocoon produced 
by the choked-jet at its shock breakout are estimated as
\begin{equation}
    \tau_{\text{d}}(z_{\text{ch,bre}})=1/\beta_{\text{ch,bre}},
\end{equation}
\begin{equation}
r_{\text{ch,bre}}=r_{\text{c}}(t_{\text{ch}})+z_{\text{ch,bre}}-z_{\text{h}}(t_{\text{ch}}),
\end{equation}
and
\begin{equation}
   \beta_{\text{ch,bre}}=\left(\frac{2 E_{\text{ch}}}
   {M_{\text{ch,bre}}c^2}\right)^{\frac{1}{2}},
\end{equation}
where $M_{\text{ch,bre}}$ represents the total mass swept by the system since the 
jet launching, ignoring the negligible mass of jet material, which is given by
\begin{equation}
M_{\text{ch,bre}}= 
\pi r_{\text{ch,bre}}^2\int_{0}^{z_{\text{ch,bre}}} \rho(z')\text{d}z'.
\end{equation}
The emission properties of the choked-jet 
cocoon can be derived directly in a manner analogous to that described in 
Section \ref{section-disk-cocoon}, by substituting $\beta_{\text{c}}$, 
$r_{\text{c}}$, $E_{\text{c}}$, $M_{\text{c}}$ with $\beta_{\text{ch,bre}}$, 
$r_{\text{ch,bre}}$, $E_{\text{ch}}$, $M_{\text{ch,bre}}$, respectively.

In the early stage after the jet being choked, the cocoon is
still clearly divided into the disk component and the jet 
component. During the overall cocoon expansion, the jet-cocoon
undergoes both a continuous transfer of its internal energy to 
the newly shocked ambient disk material and a continuous 
mixing with the disk cocoon, resulting in the gradual diminishment 
of the jet-cocoon material \citep{Eisenberg22}. When the
jet is barely choked, the top of jet-cocoon can be accelerated
to large velocity after the cocoon breakout; in contrast, 
when the jet is deeply choked within the AGN disk, the 
maximum terminal velocity of the jet-cocoon material decreases
significantly after the breakout, and in extremely deep
choking cases, the jet-cocoon may almost completely 
disappear \citep{Eisenberg22, Pais23}. To estimate the maximum
velocity of the remnant jet-cocoon, we assume that 
during the cocoon expansion after jet choking, the energy 
distribution of the jet-cocoon remains
$\text{d}E/\text{d}\,\log(\gamma_{\text{cj}} \beta_{\text{cj}})
=f_{\Gamma \beta} E_{\text{cj}}$. Furthermore, we consider that
the jet-cocoon transfers its energy to the freshly shocked disk 
materials starting from its upper part with higher terminal 
velocity. At jet choking, the height of jet-cocoon is 
$z_{\text{h}}(t_{\text{ch}})$, and its energy is $E_{\text{ch}}/2$.
When the cocoon breaks out, the jet-cocoon has swept
up disk materials from $z_{\text{h}}(t_{\text{ch}})$ to
$z_{\text{ch,bre}}$, resulting in approximately a 
$\left[1-z_{\text{h}}(t_{\text{ch}})/z_{\text{ch,bre}}\right]$ 
fraction of the jet-cocoon energy being dissipated. 
Setting the maximum terminal Lorentz factor of the 
energy-loss-free jet-cocoon to $10$, the residual 
energy determines the cut-off velocity as follows:
\begin{equation}
\frac{\log(\gamma_{\text{cut}} \beta_{\text{cut}})-\log\beta_{\text{ch,bre}}}
{\log 10-\log\beta_{\text{ch,bre}}}\simeq\frac{z_{\text{h}}(t_{\text{ch}})}{z_{\text{ch,bre}}},
\quad \gamma_{\text{cut}} \beta_{\text{cut}}= 
10^{\frac{z_{\text{h}}(t_{\text{ch}})}{z_{\text{ch,bre}}}}
\beta_{\text{ch,bre}}^{1-\frac{z_{\text{h}}(t_{\text{ch}})}{z_{\text{ch,bre}}}}.
\end{equation}
After understanding the properties of the remnant 
jet-cocoon, we can calculate its emission by applying 
the method detailed in Section \ref{section-jet-cocoon}.

\section{Observational properties}
\label{section4}
\subsection{Emission evolution}
\begin{figure*}
	\begin{center}
		\includegraphics[width=0.4\textwidth]{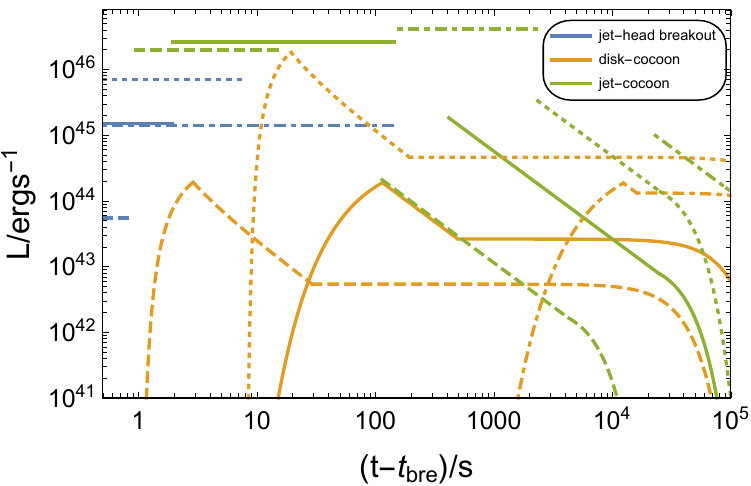}
		\quad
		\includegraphics[width=0.4\textwidth]{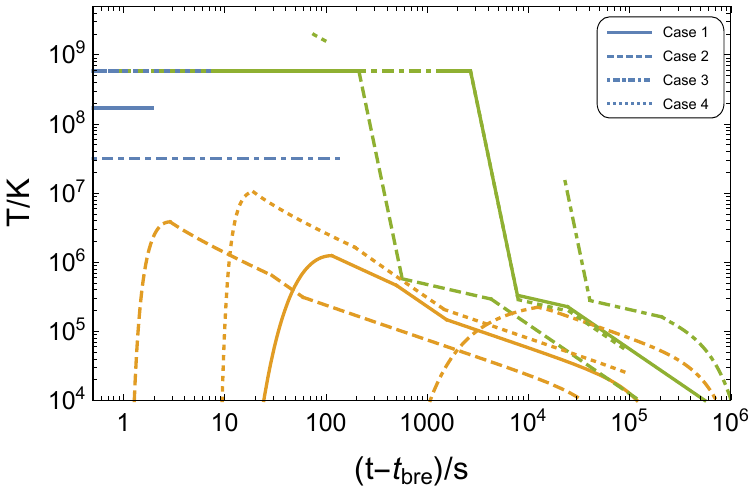}
	\end{center}
	\caption{The evolution of bolometric luminosity and 
		temperature for thermal emissions in cases 1, 2, 3, 
		and 4, varying SMBH mass, jet location, and jet power. 
		For thermal emissions, the contributions from 
		the jet-head breakout component, the disk-cocoon 
		component, and the jet-cocoon component are presented 
		separately.}
	\label{Fig:example}
\end{figure*}

\begin{figure*}
	\begin{center}
		\includegraphics[width=0.4\textwidth]{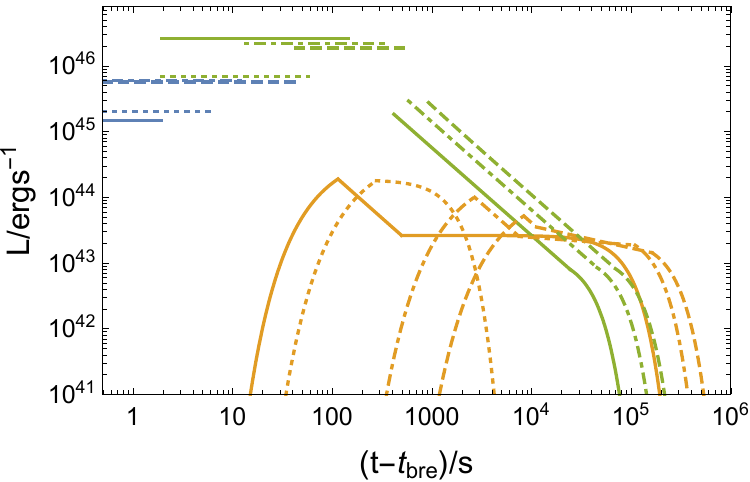}
		\quad
		\includegraphics[width=0.4\textwidth]{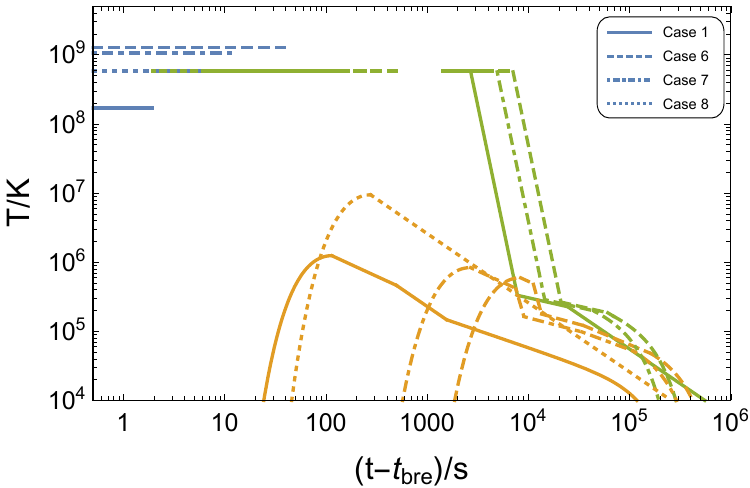}
	\end{center}
	\caption{Same as Figure \ref{Fig:example}, but 
		illustrating different cases featuring various disk 
		vertical density profiles and disk models.}
	\label{Fig:rho}
\end{figure*}

On account of the distinct properties of the jet-head,
disk-cocoon, and jet-cocoon at the jet-cocoon system breakout, 
the thermal emissions they produce exhibit significantly 
differences. Table \ref{Table1} presents a variety of 
representative cases regarding the jet launching and propagating 
in the AGN disk, of which the evolution of the bolometric 
luminosity and temperature for the corresponding thermal emission 
is shown in Figures \ref{Fig:example} and \ref{Fig:rho}.

For jet-head breakout emission, the thin width 
$d_{\text{bre}}/H_{\text{d}}\ll1$ 
and the high velocity $\beta_{\text{h}}>0.1$ of the breakout shell 
lead to a high luminosity of emission yet an extremely short 
emission duration. Meanwhile, the shell is out of thermal 
equilibrium, resulting in a high emission temperature in the 
X-ray band. Note that, for simplicity, we neglect the detailed 
temporal evolution of the emission in Figures \ref{Fig:example} 
and \ref{Fig:rho}, acknowledging that actually it is a flare that 
initially rises and subsequently declines. For disk-cocoon 
emission, the luminosity evolution exhibits a early rapid rise  
followed by a normal decline, originating from the disk-cocoon 
shock breakout process, succeeded by a phase of approximate
plateau, which results from photons being trapped in the adiabatic
expansion materials, and finally a sharp decline as
most photons have diffused out. The evolution of the observed 
temperature is relatively straightforward, showing a rapid rise
followed by a shallow decline, with the emission remaining 
essentially in thermal equilibrium. For jet-cocoon emission, the 
early constant luminosity and temperature shown in 
Figures \ref{Fig:example} and \ref{Fig:rho} are artificial,
which corresponds to a flare of photons escaping from the 
inefficiently accelerated dilute material at the top of the
jet-cocoon, carrying almost its entire internal energy. The
luminosity evolution of the Newtonian jet-cocoon material 
produced emission shows a normal decline attributed to the 
distribution of energy, mass, and velocity within the jet-cocoon, 
followed by a sharp decline as the entire jet-cocoon becomes 
radiative. The temperature evolution of the emission 
demonstrates an extremely sharp decline due to the sensitive 
Comptonized radiation, succeeded by a shallow decline in thermal
equilibrium radiation. Note that the emission would transition 
continuously from the relativistic phase to the Newtonian phase. 
However, since the transition phase is difficult to calculate 
precisely \citep{Nakar17}, we neglect this phase in our analysis.
 
Next, we investigate the effects of environmental factors and jet 
properties on the emissions. 
Comparing case 1, 2, and 3, the generation time of 
emission is strongly influenced by the height of AGN disk, 
which is because the jet-cocoon system must break out of the AGN disk 
to produce observable emissions. In contrast to other components, 
the luminosity of Newtonian jet-cocoon emission demonstrates a 
stronger dependence on the properties of the AGN disk, as 
in Equation (\ref{Lcjb}), both $\beta_{\text{cj}}$ and 
$V_{\text{cj},\beta}$ are significantly modulated by variations in disk 
density and height. For all emission components, the earlier 
they are produced, the higher their peak observed temperature, unless 
pair production suppresses the temperature at approximately 
$50\,\text{keV}$ for the early Newtonian jet-cocoon component emission.
As demonstrated by cases 1, 4 and 5, an increase in jet power results in 
an earlier breakout, characterized by a faster jet-head and cocoon.
Consequently, emissions are generated earlier, with all emission 
components becoming more powerful and achieving higher temperatures.
Furthermore, distinct disk vertical density profiles and disk 
models can substantially influence the emission properties even
for the same jet. In cases
6 and 7, the reacceleration of the jet-head leads to a brighter 
shock breakout emission with higher observed temperature and longer 
duration. The emission properties of the jet-cocoon remain largely 
unchanged. Conversely, due to having swept up more disk gas, 
the disk-cocoon becomes radiative later, both its emission luminosity 
and peak temperature are slightly lower compared to those of a more 
compact uniform disk; these similar properties are also manifested 
in supernova explosions in AGN disks \citep{Grishin21}. In case 8,
the disk adopts the model described in \cite{Thompson05}, representing 
a lower-density disk. Consequently, the jet-cocoon becomes sufficiently 
dilute, enabling photons to escape from the entire structure and 
thereby producing emission immediately following the system breakout,
which results in the absence of relativistic or Newtonian jet-cocoon 
emission. Moreover, the low density of the disk-cocoon, combined with 
its high expansion velocity, causes a relatively rapid luminosity 
evolution and a higher peak observed temperature.

\begin{table}[h]
	\caption{Cases of jet launching and propagating in 
		the AGN disk. $M_{\text{SMBH}}$ (in $M_{\odot}$) is the mass 
		of the central SMBH; $R_{\text{j}}$ (in $R_{\text{g}}$) is the 
		jet location in the AGN disk; $\rho_{0}$ 
		(in $\text{g\,cm}^{-3}$) and $H_{\text{d}}$ (in $\text{cm}$) 
		are the mid-plane density and half-thickness of the AGN 
		disk; $\rho_{\text{z}}$ represents the disk vertical density 
		profile (Uni for uniform, Iso for isothermal, and Poly for
		polytropic) and disk model (TQM for \cite{Thompson05} model); 
		$L_{\text{j}}$ (in $\text{erg\,s}^{-1}$)
		is the jet power, where all the jets are set to be continuous. 
		The properties of the jet-cocoon system at its breakout are 
		given: $t_{\text{bre}}$ (in $\text{s}$) is the breakout time 
		starting from the jet launching, $d_{\text{bre}}$ (in $H_{\text{d}}$)
		is the width of the breakout shell, $\beta_{\text{h}}$ and 
		$\beta_{\text{c}}$ are the velocities of the jet-head and the
		disk-cocoon, $\theta_{\text{h}}$ and $\theta_{\text{c}}$ are
		the opening angle of the jet-head and the disk-cocoon.}
	\label{Table1}
	\centering
	\begin{tabular}{ccccccccccccc}
		\hline	
		\hline	
		case & $M_{\text{SMBH}}$ & $R_{\text{j}}$ & 
		$\rho_{0}$ & $H_{\text{d}}$ & $\rho_{\text{z}}$ &
		$L_{\text{j}}$ & $t_{\text{bre}}$ & $d_{\text{bre}}$ & $\beta_{\text{h}}$ &
		$\beta_{\text{c}}$ & $\theta_{\text{h}}$ & $\theta_{\text{c}}$ \\
		\hline
		1 & $10^{7}$ & $10^{4}$ & $4.1 \times 10^{-10}$ & $4.3 \times 10^{13}$ &
		Uni & $10^{46}$ & $3.2 \times 10^{3}$ &
		$4.7 \times 10^{-4}$ & 0.35 & 0.046 & 0.023 & 0.17 \\
		\hline
		2 & $10^{7}$ & $10^{3}$ & $1.3 \times 10^{-8}$ & $4.3 \times 10^{12}$ &
		Uni & $10^{46}$ & 272.6 &
		$1.2 \times 10^{-4}$ & 0.42 & 0.062 & 0.031 & 0.20 \\
		\hline
		3 & $10^{8}$ & $10^{4}$ & $9.8 \times 10^{-12}$ & $7.0 \times 10^{14}$ &
		Uni & $10^{46}$ & $6.8 \times 10^{4}$ &
		$1.7 \times 10^{-3}$ & 0.25 & 0.029 & 0.015 & 0.14 \\
		\hline
		4 & $10^{7}$ & $10^{4}$ & $4.1 \times 10^{-10}$ & $4.3 \times 10^{13}$ &
		Uni & $10^{48}$ & $2.0 \times 10^{3}$ &
		$2.6 \times 10^{-4}$ & 0.63 & 0.14 & 0.072 & 0.31 \\
		\hline
		5 & $10^{7}$ & $10^{4}$ & $4.1 \times 10^{-10}$ & $4.3 \times 10^{13}$ &
		Uni & $L_{\text{b,AGN}}$ & $4.5 \times 10^{3}$ &
		$6.9 \times 10^{-4}$ & 0.24 & 0.027 & 0.014 & 0.14 \\
		\hline
		6 & $10^{7}$ & $10^{4}$ & $4.1 \times 10^{-10}$ & $4.3 \times 10^{13}$ &
		Iso & $10^{46}$ & $1.0 \times 10^{4}$ &
		$7.7 \times 10^{-3}$ & 0.89 & 0.032 & 0.015 & 0.10 \\
		\hline
		7 & $10^{7}$ & $10^{4}$ & $4.1 \times 10^{-10}$ & $4.3 \times 10^{13}$ &
		Poly & $10^{46}$ & $6.6 \times 10^{3}$ &
		$8.2 \times 10^{-2}$ & 0.83 & 0.038 & 0.017 & 0.12 \\
		\hline
		8 & $10^{7}$ & $10^{4}$ & $1.2 \times 10^{-11}$ & $1.7 \times 10^{13}$ &
		TQM & $10^{46}$ & 752.0 &
		$2.1 \times 10^{-2}$ & 0.67 & 0.17 & 0.089 & 0.34 \\
		\hline
	\end{tabular}
\end{table}

\subsection{Detectability}
\label{section4-2}
\begin{figure*}
	\begin{center}
		\includegraphics[width=0.32\textwidth]{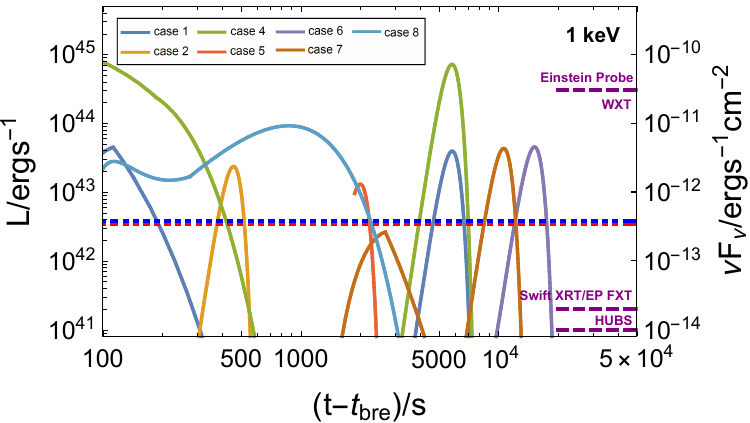}
		\includegraphics[width=0.32\textwidth]{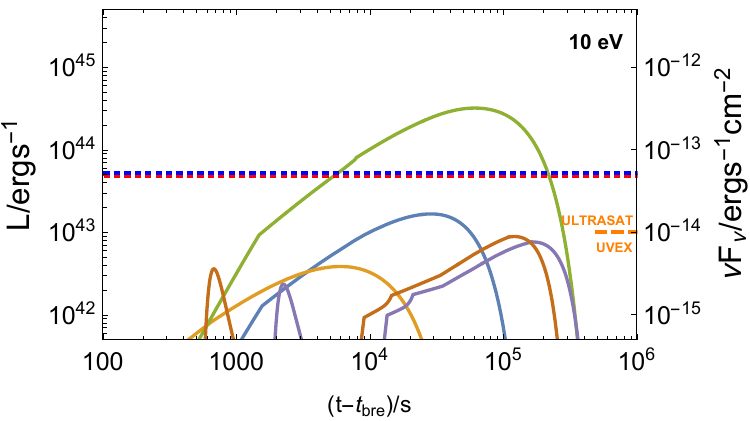}
		\includegraphics[width=0.325\textwidth]{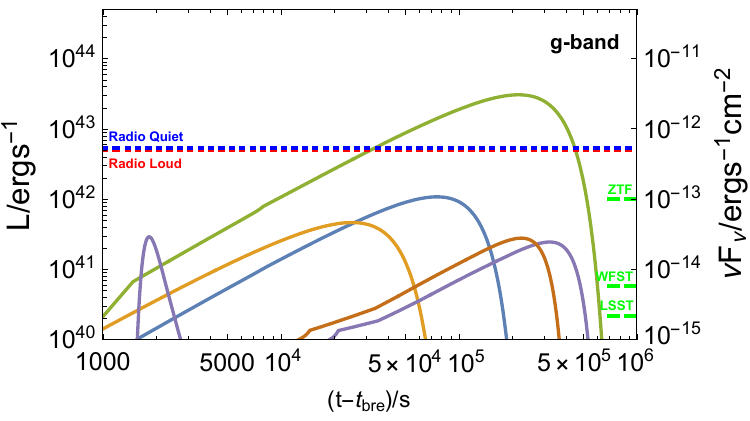}
	\end{center}
	\caption{Light curves in soft X-ray ($1\,\text{keV}$), 
		UV ($10\,\text{eV}$), and optical (g-band) band for various cases. 
		The red and blue dotted lines in each panel represent 
		the typical luminosity of radio-loud and radio-quiet AGN with 
		$M_{\text{SMBH}}=10^7M_{\odot}$ in these bands, which are
		calculated by adopting the mean spectral energy distribution
		of quasars \citep{Shang11}. To test detectability, the left and 
		right panels assume a source luminosity distance of 
		$d_{\text{L}}=300\,\text{Mpc}$, 
		while the middle panel assumes $d_{\text{L}}=3\,\text{Gpc}$.
		The dashed purple lines mark the sensitivity limit of EP WXT for 
		$1\,\rm{ks}$ exposure time, Swift XRT, EP FXT, and HUBS for 
		$10\,\rm{ks}$ exposure time (note that the sensitivity limits of Chandra 
		and XMM-Newton are $\sim 10^{-15}\,\rm{erg}\,\rm{s}^{-1}\,\rm{cm}^{-2}$
		for $10\,\rm{ks}$ exposure time, 
		below the lowest vertical value of the left panel); the 
		dashed orange lines indicate the 
		sensitivity limit of ULTRASAT and UVEX for $\sim 1 \,\rm{ks}$ 
		exposure time; the green lines denote the sensitivity limit of
		ZTF, WFST, and LSST for $30\,\rm{s}$ exposure time.
		}
	\label{Fig:obs}
\end{figure*}

\begin{figure*}
	\begin{center}
		\includegraphics[width=0.5\textwidth]{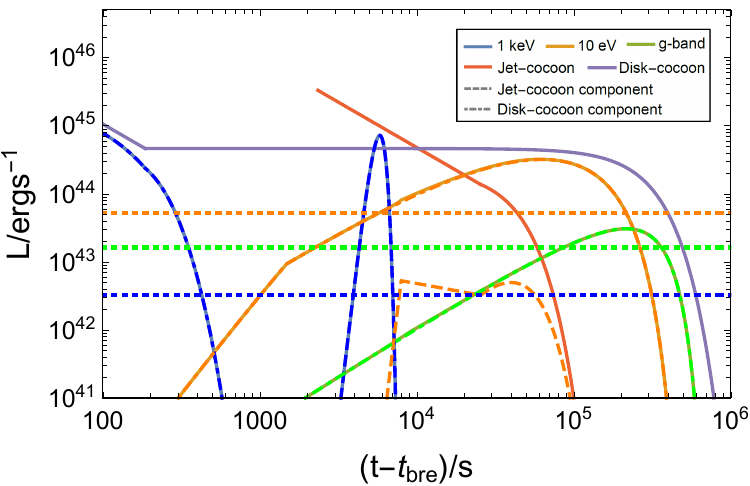}
	\end{center}
	\caption{Disk-cocoon and jet-cocoon contributions to 
		the specific-band radiation in case 4. The purple and
		orange lines represent the bolometric luminosity of the 
		disk-cocoon emission and jet-cocoon emission, respectively. 
		The blue, golden, and green lines depict the overall light 
		curves for X-ray, UV, and optical band, superimposed by the
		disk-cocoon component shown as a dot-dash line and 
		the jet-cocoon component as a dashed line. Additionally,
		the blue, orange, and green dotted lines illustrate the 
		radiation of the background AGN.}
	\label{Fig:component}
\end{figure*}

\begin{figure*}
	\begin{center}
		\includegraphics[width=0.3\textwidth]{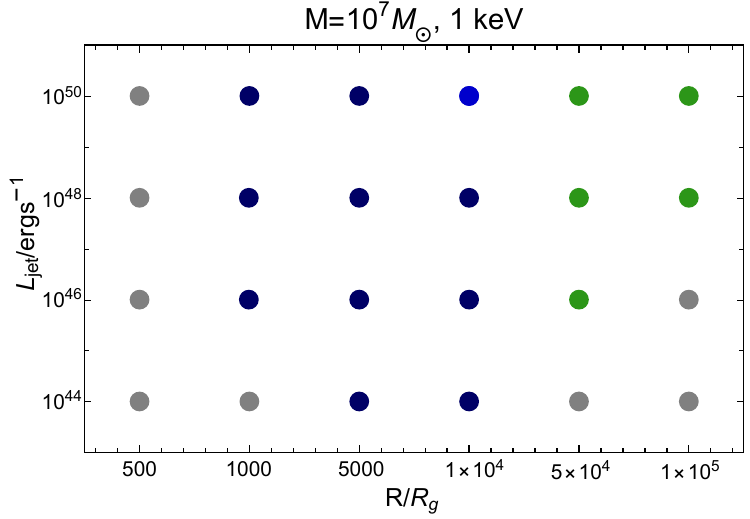}
		\includegraphics[width=0.3\textwidth]{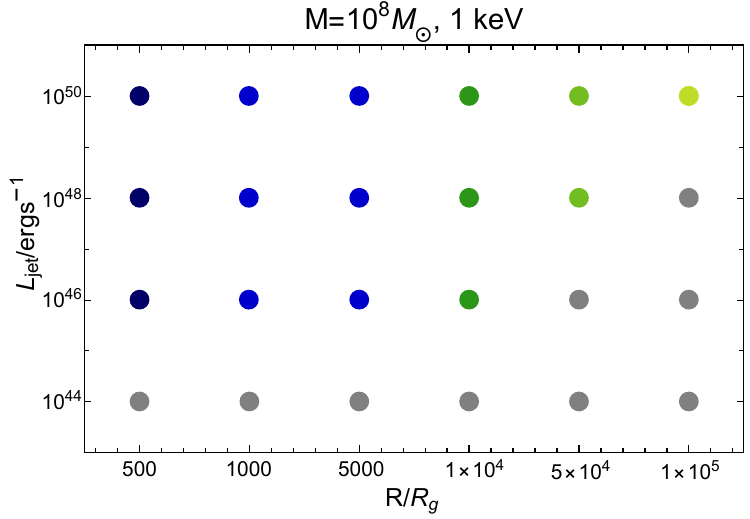}
		\includegraphics[width=0.3\textwidth]{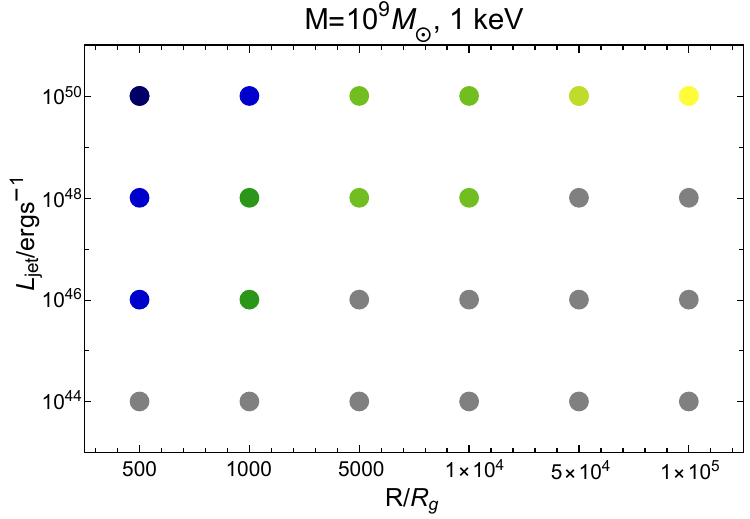}
		\includegraphics[width=0.3\textwidth]{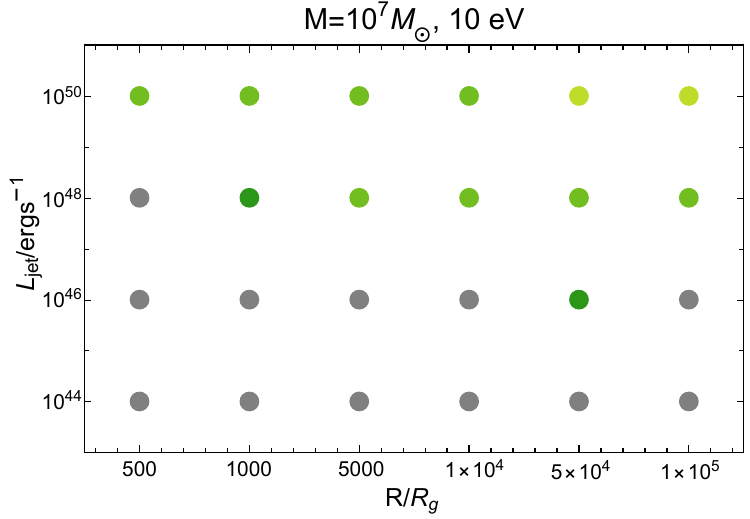}
		\includegraphics[width=0.3\textwidth]{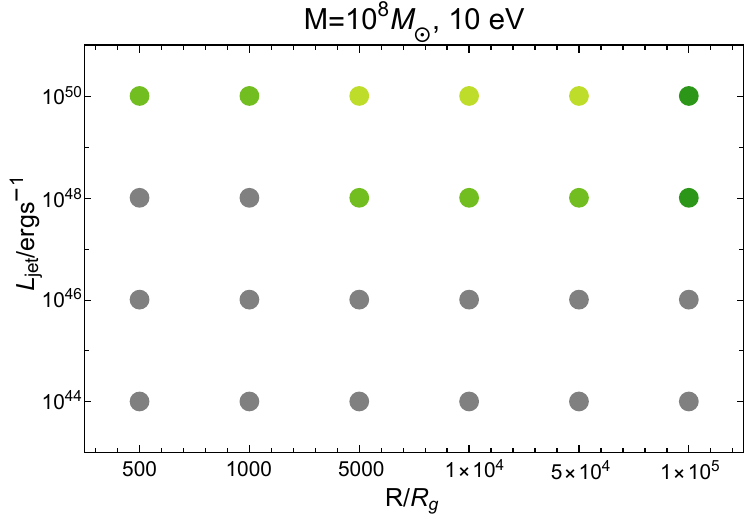}
		\includegraphics[width=0.3\textwidth]{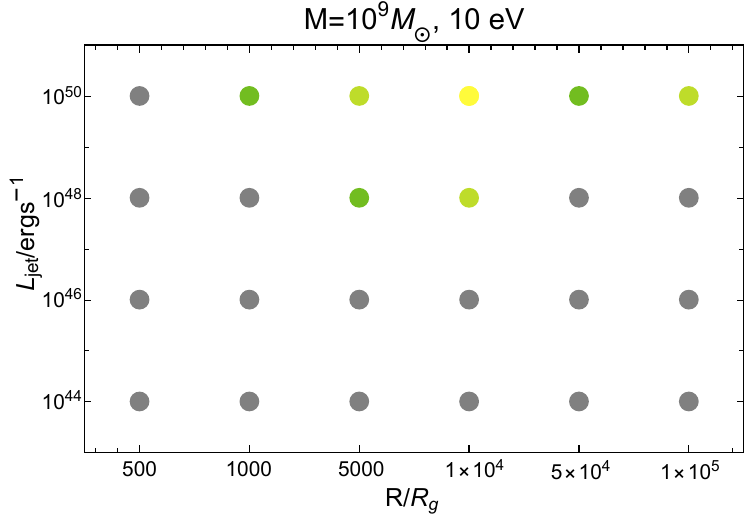}
		\includegraphics[width=0.3\textwidth]{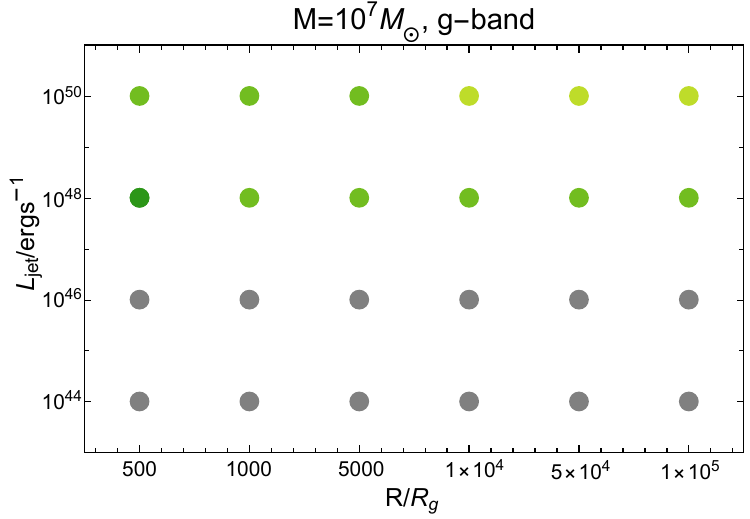}
		\includegraphics[width=0.3\textwidth]{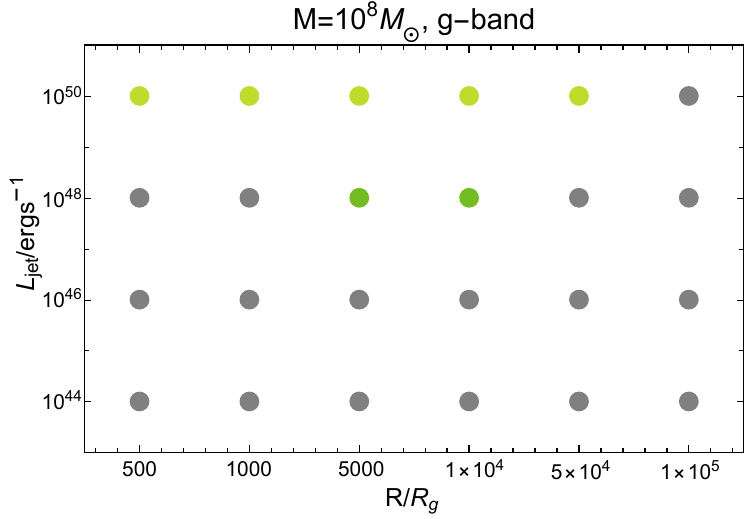}
		\includegraphics[width=0.3\textwidth]{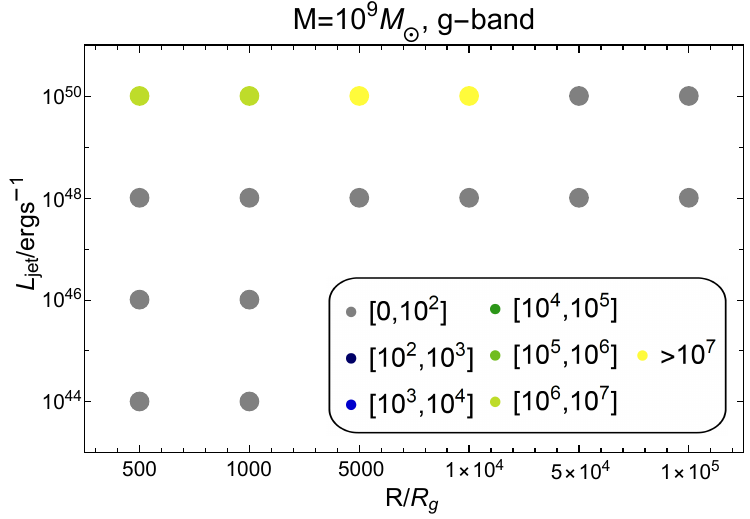}
	\end{center}
	\caption{Duration of the observable flares in X-ray 
		(first row), UV (second row), and optical (last row)
		bands produced by jets with varying powers embedded
		within different radius of AGN disks with central
		SMBH mass of $10^7 M_{\odot}$ (first column), 
		$10^8 M_{\odot}$ (second column), and $10^9 M_{\odot}$
		(last column). Dots in various colors represent 
		the range of duration that the jet-cocoon system produced
		emission brighter than the AGN background radiation.}
	\label{Fig:bands}
\end{figure*}

We now investigate the detectability of these thermal emissions, with
particular emphasis on the soft X-ray ($1\,\text{keV}$), UV ($10\,\text{eV}$), 
and optical (g-band) bands. If brighter than the AGN background radiation,
the soft X-ray flare can be captured by telescopes such as Chandra 
\citep{Weisskopf00}, XMM-Newton \citep{Jansen01}, Swift XRT \citep{Burrows05}, 
Einstein Probe (EP) \citep{Yuan22}, and HUBS \citep{Zhang22}; 
the UV flare can be observed by
facilities like ULTRASAT \citep{Sagiv14}, and UVEX \citep{Kulkarni21};
the optical can be detected by telescopes such as LSST \citep{LSST09}, 
ZTF \citep{Bellm14}, and WFST \citep{Lei23}.

\subsubsection{Soft X-ray}
Light curves in the soft X-ray band for various cases are presented in 
Figure \ref{Fig:obs}. As the AGN background radiation in this band is
relatively weak, the X-ray flares produced by jet-cocoon systems are 
distinguishable in these cases. The elevated luminosity enhances the 
observed flux surpassing the sensitivity thresholds of various soft X-ray 
telescopes (as exemplified in the left panel of Figure \ref{Fig:obs}), 
thus enabling the detection of the flare. Many flares exhibit a profile 
that rises rapidly and then declines sharply, originating from the 
Newtonian jet-cocoon component as illustrated in Figure \ref{Fig:component}.
Since these temperatures follow the same Compton evolution, the
profiles of the flares are homologous. Meanwhile, in the early stage
of cases 1, 4, (7) and 8, additional short-duration flares are observed, 
which are produced by the fast disk-cocoon component as illustrated in 
Figure \ref{Fig:component}. Consequently, some soft X-ray flares 
displays a double-peak characteristics, with the first peak showing 
a sharper profile. Figure \ref{Fig:bands} depicts the duration of the 
observable soft X-ray flare across various parameter space regions.
We find that the X-ray flares are generally observable, 
as their peak luminosities
typically exceed the AGN background levels. However, their detectability 
additionally depends on both the luminosity distance of the source and 
telescope sensitivity. Exceptions occur when the 
jet power is substantially low, the SMBH mass is exceedingly high or the
jet launching location is situated in the very inner region of the AGN disk.
In most parameter spaces, the flare persists for $<O(10^5)\,\text{s}$ or 
a few days, signifying a short-lived event. Additionally, the flare 
duration increases with an increase in SMBH mass or AGN disk radius, 
attributable to the larger energy and volume of the jet-cocoon materials.

\subsubsection{UV and Optical}
The UV flares illustrated in the cases presented in Figure 
\ref{Fig:obs} are generally unobservable, as their peak luminosities 
are dimmer than the AGN background. The lack of detectability 
can be attributed to two main factors. First, the strong AGN 
radiation in the UV band masks weaker signals, making them 
difficult to distinguish. Second, when the observed temperature 
shifts into the UV band, both the disk-cocoon and the jet-cocoon, 
which responsible for the double-peak feature in cases 6 and 7, 
have undergone significant adiabatic expansion, leading to less 
energetic emissions, thereby reducing the detectability of the UV 
flares. Even so, an observable UV flare can be generated by the 
jet-cocoon system when the jet possesses sufficient power. For
instance, as shown in Figure \ref{Fig:obs}, a jet with a 
power of $10^{48}\,\text{erg\,s}^{-1}$ in case 4 can produce 
a strong UV flare, which is primarily contributed by the 
disk-cocoon component. Owing to the high sensitivity of 
the advanced UV telescopes, even distant flares have the potential 
to be  detected. The properties of the UV flare are
systematically illustrated in Figure \ref{Fig:bands}. Only jets
with power $>10^{46}\,\text{erg\,s}^{-1}$ can produce 
observable UV flares, and the required power increases as
the SMBH mass grows. An optimistic point is that the observable 
flare are relatively long-lasting, with duration ranging from a few 
days to several tens of days. However, for jets with extremely high 
power, e.g., $10^{50}\,\text{erg\,s}^{-1}$, their central engine 
must operate for at least $t_{\text{j,cri}}$ (see Figure \ref{Fig:tj}) 
to support a successful jet breakout. It is quite challenging to 
satisfy this condition in astrophysical systems, and when 
$t_{\text{j}}<t_{\text{j,cri}}$, the choked jet-cocoon system would 
produce a dimmer flare (see below), potentially rendering it 
unobservable.

The properties of the optical flares resemble those of the UV flares.
However, they exhibit weaker power and a more delayed peak time 
due to the lower corresponding emission temperature. As a result,
optical flares are more difficult to observe, despite the observable 
flares lasting for a longer duration, as shown in Figure \ref{Fig:bands}.
Nonetheless, under favorable system parameter conditions, 
these flares can be captured by various telescopes, as demonstrated in 
Figure \ref{Fig:obs}.

\subsection{Choked jet emission}
\begin{table}[h]
\caption{Cases of choked jet in the AGN disk. The first five
parameters have the same meaning as those in Table \ref{Table1}.
$t_{\text{j}}$ (in $\text{s}$) is the duration of the jet, 
where the first row is the continuous jet serving as a reference; 
$t_{\text{ch}}$ (in $\text{s}$) is the jet choking time; 
$t_{\text{bre}}$ (in $\text{s}$) is the breakout time of the
successful jet or the jet-choked cocoon; 
$z_{\text{ch}}$ (in $H_{\text{d}}$) is the height at which the jet 
becomes choked; $d_{\text{ch,bre}}$ (in $H_{\text{d}}$) and 
$\beta_{\text{ch,bre}}$ are the width and velocity of the cocoon
breakout shell; $\theta_{\text{ch,bre}}$ is the opening angle of 
the jet-choked cocoon at its breakout; and $\beta_{\text{cut}}$ is 
the maximum terminal velocity of the jet-choked jet-cocoon.
The AGN disk adopts the \cite{Sirko03} model, assuming a uniform
vertical density profile.}
\label{Table2}
\centering
\begin{tabular}{cccccccccccccc}
\hline	
\hline	
case & $M_{\text{SMBH}}$ & $R_{\text{j}}$ & $\rho_{0}$ & $H_{\text{d}}$ & 
$L_{\text{j}}$ & $t_{\text{j}}$ & $t_{\text{ch}}$ & $t_{\text{bre}}$ & 
$z_{\text{ch}}$& $d_{\text{ch,bre}}$ & $\beta_{\text{ch,bre}}$ & 
$\theta_{\text{ch,bre}}$ & $\beta_{\text{cut}}$ \\
\hline
9 & $10^{8}$ & $10^{3}$ & $2.3 \times 10^{-9}$ & 
$3.5 \times 10^{13}$ & $10^{50}$ & $-$ & $-$ & 
$1.3 \times 10^{3}$ & $-$ & $4.6 \times 10^{-5}$ & 
0.29 & 0.44 & $-$ \\
\hline
10 & $10^{8}$ & $10^{3}$ & $2.3 \times 10^{-9}$ & 
$3.5 \times 10^{13}$ & $10^{50}$ & 1 & 95 & 
$5.0 \times 10^{4}$ & 0.08 & $1.7 \times 10^{-3}$ & 
0.022 & 0.98 & 0.035 \\
\hline
11 & $10^{8}$ & $10^{3}$ & $2.3 \times 10^{-9}$ & 
$3.5 \times 10^{13}$ & $10^{50}$ & 10 & 307.9 & 
$1.3 \times 10^{4}$ & 0.25 & $5.3 \times 10^{-4}$ & 
0.067 & 0.89 & 0.23 \\
\hline
12 & $10^{8}$ & $10^{3}$ & $2.3 \times 10^{-9}$ & 
$3.5 \times 10^{13}$ & $10^{50}$ & 100 & 
$1.0 \times 10^{3}$ & $1.7 \times 10^{3}$ & 
0.80 & $1.1 \times 10^{-4}$ & 0.23 & 0.57 & 0.98 \\
\hline
\end{tabular}
\end{table}

Table \ref{Table2} presents four cases for investigating
the jet choking of $t_{\text{ch}}<t_{\text{bre}}$, 
where case 9 represents a successful breakout jet and
serves as a control example, while case 10 denotes a 
deeply choked jet, case 11 indicates a moderately choked jet, 
and case 12 corresponds to a barely choked jet. For jets
with the same power, a shorter central engine active duration
leads to a deeper choking, as the jet tail at engine 
shut-off would cover a shorter distance to catch up 
with its head. Both the velocity of the jet-choked cocoon 
at its breakout, $\beta_{\text{ch,bre}}$, and the cut-off terminal 
velocity of the jet-choked jet-cocoon, $\beta_{\text{cut}}$, decrease
as $z_{\text{ch}}$ decreases, indicating that the deeper choking
leads to a less powerful cocoon, owing to an increased opening angle, 
a correspondingly larger cocoon volume and mass, and consequently
a reduction in cocoon energy density.

\begin{figure*}
	\begin{center}
		\includegraphics[width=0.4\textwidth]{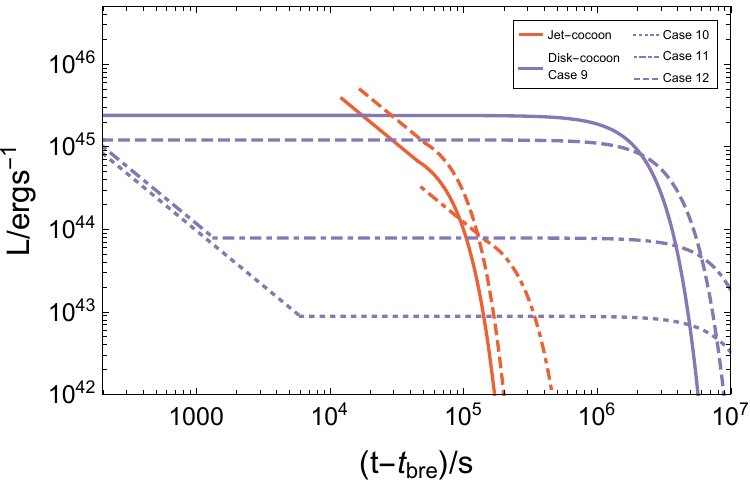}
		\includegraphics[width=0.46\textwidth]{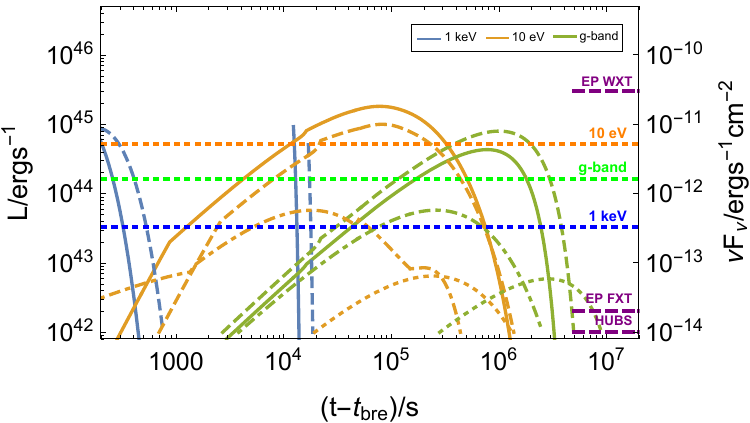}
	\end{center}
	\caption{The evolution of bolometric luminosity for 
		thermal emissions from the disk-cocoon and 
		jet-cocoon components in cases 9, 10, 11, and 12, 
		as well as the corresponding light curves in the
		X-ray, UV, and optical bands. The blue, 
		orange, and green dotted lines illustrate the 
		radiation of the background AGN. For 
		a source luminosity distance of 
		$d_{\text{L}}=300\,\text{Mpc}$, the dashed purple 
		lines are identical to those in Figure 
		\ref{Fig:obs}.}
	\label{Fig:choked}
\end{figure*}

The emission produced by the jet-choked cocoon system is
weaker compared to that of a successful jet-cocoon system.
As shown in Figure \ref{Fig:choked}, the soft X-ray, UV,
and optical flare emitted in case 9 are more luminous than
the AGN background radiation. Therefore, this successful 
jet breakout event represents a multi-wavelength phenomenon.
When the jet is barely choked in case 12, the overall 
properties of the cocoon have not changed significantly.
A slightly longer breakout time and a modestly increased cocoon 
volume, together with a moderate decrease in cocoon velocity, 
result in a subtle deviation in the luminosity and temperature 
evolution of both cocoon components. Consequently, the flares
are only slightly higher or lower than those associated with 
successful jets, rendering all flares in this case observable.
However, the overall emission luminosity of the cocoon 
decreases substantially when the choking height is significantly 
deeper, as the cocoon becomes much weaker. As a result, 
in cases 10 and 11, the UV and optical flares produced are dimmer 
than those in cases 9 and 12, rendering them undetectable. 
Meanwhile, for a more deeply choked jet system, the jet-cocoon 
materials have been significantly weakened at the cocoon breakout. 
In case 11, the luminosity of the jet-cocoon emission decreases 
significantly, and since $\beta_{\text{cut}}$ cannot result in the 
shock temperature reaching the X-ray band, no soft X-ray flare
is generated. Moreover, in case 10, the choking is exceptionally 
deep, causing the freshly shocked disk materials to entirely 
drain the energy of the jet-cocoon, leading to its complete 
dissipation.

\subsection{Specific astrophysical events}
The potential existence of EM counterparts associated 
with GW events constitutes a distinctive property of 
binary black holes mergers occurring within the AGN disk. 
Indeed, several candidate optical flares have been 
reported to in association with LIGO GW events 
\citep{Graham20, Graham23}, where the host AGNs harbor 
massive SMBH with masses of $\sim 10^8-10^9M_{\odot}$. 
If jets are the energy sources for radiation 
\citep[e.g.][]{Tagawa23b, Rodrguez24, ChenK24}, and these
flares are indeed the thermal emissions produced by the 
jet-cocoon system investigated in this work, then the jet 
powers are required to $\geqslant10^{48}\,\text{erg\,s}^{-1}$ 
(as shown in Figure \ref{Fig:bands}), 
and the jets must persist for a sufficient duration 
$\gtrsim t_{\text{j,cri}}$ to ensure a successful breakout. 
These constraints on jet power and duration can facilitate 
the testing and investigation of the jet-launching mechanism of 
merger-remnant black holes. Meanwhile, the observation of 
optical flares is often accompanied by brighter soft X-rays 
and UV emissions. This suggests that merger-remnant black 
holes are multi-wavelength sources, and therefore 
multi-wavelength observation can serve as a crucial method to 
examine whether the candidate optical flares are the actual 
EM counterparts.

As a well-known EM counterpart for GW event, short GRBs 
arise from jets driven by the remnants of binary 
neutron star mergers or neutron star-black hole mergers. 
As short GRBs typically last less than a few seconds, the
jets produced by these mergers within AGN disks are more 
likely to be choked unless the jets possess exceptional 
power and are launched at an extremely small inner disk radius,
e.g., a location within a few hundred $R_{\text{g}}$ around 
a less massive SMBH $\sim 10^7 M_{\odot}$, as illustrated in 
Figure \ref{Fig:tj}. For short GRB jets that successfully 
break out of AGN disks, normal prompt emissions and 
afterglows can be produced \citep{Perna21a}. 
Additionally, the cocoon material can contribute an 
extra early thermal emission component, which manifests 
as a UV flare and an optical flare lasting several days, 
while the X-ray flare is difficult to observe due to its
short duration $\lesssim O(10^3)\,\text{s}$. Conversely, when
short GRB jets are choked in AGN harboring massive SMBH
$\gtrsim 10^8M_{\odot}$, no thermal flare are observable,
let alone the absent normal prompt emission and 
afterglow. Only in cases where the jets are barely choked
do thermal flares become observable, serving as the sole 
EM signals of the system when the subsequent non-thermal
emission from the interaction between the cocoon material 
and the ambient gas beyond the AGN disk is weak. 

The jet-driven long GRBs, which originate from the 
collapse of massive stars, constitute a class of strong 
transient sources. Stars can form in large numbers 
within the outer self-gravity unstable regions of AGN disks 
through fragmentation processes \citep{Goodman03}.
However, whether these outer stars can migrate to the 
inner regions of AGN disks depends crucially on 
the still-debated efficiency of migration mechanisms
\citep[e.g.][]{Bellovary16, Grishin24, Wu24}. Meanwhile, 
the dense inner regions of AGN disks are capable 
of capturing a substantial number of obliquely orbiting 
stars from the nuclear star cluster \citep[]{Fabj20}. 
Moreover, the accretion processes of stars in the 
dense environments of these inner disk regions remain 
poorly understood by both theoretical models and numerical 
simulations \citep[e.g.][]{ChenYX24b, Dittmann25, Fabj25}.
To date, the comprehensive distribution and evolution of 
the embedded stellar population have not yet been 
thoroughly studied. Consequently, the specific statistical 
properties of AGN-disk-assisted long GRB jets remain 
inadequately constrained, particularly concerning their 
launch locations. Given that long GRBs feature more powerful 
and sustained jets, if they indeed occur within optically 
thick inner AGN disks, the generation of observable soft X-ray, 
UV, and optical thermal flares becomes more probable.

For any other astrophysical events embedded in AGN disks 
that launch jets, the properties of thermal emissions 
can be directly determined using the methods presented in 
this work, provided that the power and duration of the 
jets are specified by the central engine launching 
mechanism. 

\section{Summary and Discussion}
\label{section5}
In this study, we have systematically investigated the 
dynamics and thermal emission properties of relativistic 
jets embedded in AGN disks. We calculated the 
evolution of the jet-cocoon system, by considering both 
successful breakout and choked jets, and 
analyzed the resulting thermal radiation from the 
jet-head shock breakout, disk-cocoon, and jet-cocoon 
components. Varying jet power, jet duration, and AGN disk 
parameters, we found that soft X-ray flares emerge 
as the most prominent observable signals, with duration 
ranging from $O(10^2)\,\text{s}$ to $O(10^5)\,\text{s}$,
frequently displaying double-peaked light curve morphology. 
UV flares produced in systems hosting ultra-powerful jets 
($L_{\text{j}}\geqslant10^{48}\,\text{erg\,s}^{-1}$) 
that either successfully break out or are barely choked can 
be observed, possessing relatively longer duration ranging 
from several days to tens of days. Observable optical flares 
are the most challenging to produce; while they potentially 
distinguishable from AGN background, their luminosity remains 
lower than those of the other two types of flares, coupled 
with protracted emergence timescales and extended duration.
Crucially, these multi-wavelength signatures establish a 
diagnostic paradigm for identifying whether the observed optical 
flares are genuine EM counterparts to binary black hole merger 
GW events within AGN disks, particularly through their 
characteristic temporal sequencing where optical flares 
consistently follow precursor UV and soft X-ray emissions.
Moreover, these thermal flares represent an essential 
component of the early observable signatures for the breaking-out 
embedded GRB jets, and may even constitute the sole detectable 
signature when the jet is choked within the AGN disk.

Our study primarily examines the thermal emissions produced
by the initially opaque jet-cocoon system during the early 
stage following the breakout of the jet and cocoon. 
Upon emergence from the AGN disk, the relativistic jet 
(if it persists) and accompanying cocoon materials interact 
with the ambient gas above the AGN disk photosphere, such 
as in the broad-line region, through collisionless shocks 
that generate non-thermal radiation. By contrast, jets 
launched from the outer regions of the AGN disk exhibit a
distinct behavior. In these optically-thin environments 
where the disk total optical depth satisfies 
$\tau_{\text{d}}(z=0)<1$, 
the shocked jet-head undergoes collisionless processes that 
directly produce non-thermal radiation, bypassing the formation 
of a cocoon structure. The detailed analysis of the system
evolution and the resulting non-thermal emissions, while beyond 
the current scope, will be studied elsewhere.

Additionally, it should be noted that we have assumed that 
the jets originate from the mid-plane of an undistributed AGN 
disk and propagate perpendicular to the disk plane. Despite 
being reasonable and representative, the assumptions are somewhat 
oversimplified. Below, we briefly discuss the implications of 
relaxing these idealized conditions. First, prior to or during 
jet launching, persistent accretion of disk materials by the 
central engine potentially alter the gas distribution along the 
jet propagation path. 
For instance, if accretion leads to significant 
large-scale gas accumulation around the central engine, the 
enhanced ambient density would exert stronger deceleration 
on the jet. Conversely, if accretion-driven outflow feedback 
dominates, a low-density cavity could form along the jet
trajectory \citep[e.g.][]{Wang21a, ChenK23, Tagawa24}, 
significantly reducing environmental resistance and enabling 
the jet to propagate more freely. Second, vertical displacement 
of the launching site from the disk midplane introduces asymmetric 
propagation distances: compared to the mid-plane case, the 
jet facing the nearer disk surface encounters a reduced 
amount of AGN disk material, while the opposing jet 
must traverse a greater disk thickness. For a successful breakout
jet, a larger propagation distance implies more energy being 
deposited into the cocoon, thereby enhancing the brightness of 
the cocoon-produced emission. Conversely, midplane-choked jets 
might achieve breakout if launched closer to the disk surface. 
Finally, jet inclination adds additional complexity: oblique 
trajectories require longer breakout paths through the disk 
medium. Consequently, successful inclined jets would 
transfer more energy to cocoon, potentially generating brighter 
thermal flares, while jets originally choked at the disk 
mid-plane would experience deeper choking.

\section{Acknowledgments}
This work was supported by  the National Natural Science Foundation of China (grant No. 12393812), the National SKA Program of China (grant No. 2020SKA0120302), and the Strategic Priority Research Program of the Chinese Academy of Sciences (grant NO. XDB0550300).

\begin{appendix}
\section{Adiabatic Evolution of Relativistic Breakout Shell}
\label{Appendix-Rel-shell}
The acceleration of a relativistic optically thick shell during the 
planar phase following its breakout can be described by 
\citep{Pan06, Nakar12}
\begin{equation}\label{gacc}
	\gamma=\gamma_{\text{h}}\left(\frac{t}{t_0}\right)^{\frac{\sqrt{3}-1}{2}},
\end{equation}
where $t_0\sim1/\kappa\rho c$. When the entire internal energy is adiabatically 
converted into kinetic energy, the shell reaches its final Lorentz factor
\citep{Johnson71, Pan06}
\begin{equation}
	\gamma_{\text{f}}=\gamma_{\text{h}}^{\sqrt{3}+1},
\end{equation}
and the acceleration ends at
\begin{equation}
	t_{\text{f}}=t_0\gamma_{\text{h}}^{3+\sqrt{3}}.
\end{equation}
The transition from planar to spherical evolution occurs when the shell's 
width matches its initial radius at the shell's radius doubling, which
is around
\begin{equation}
	t_{\text{h,s}}\simeq\frac{(\Sigma_{\text{h}}/\pi)^{\frac{1}{2}}}{c}.
\end{equation}
A critical Lorentz factor $\gamma_{\text{h,s}}$ is determined by the condition
$t_{\text{f}}=t_{\text{h,s}}$:
\begin{equation}
	\gamma_{\text{h,s}}=
	\left[\frac{(\Sigma_{\text{h}}/\pi)^{\frac{1}{2}}}{t_0 c}\right]^{\frac{1}{3+\sqrt{3}}},
\end{equation}
which can divide the shell evolution into two distinct modes. For 
$\gamma_{\text{h}}\leqslant\gamma_{\text{h,s}}$, acceleration terminates during
the planar phase and the corresponding shell final Lorentz factor is 
$\gamma_{\text{f}}$. Otherwise, acceleration ends in the spherical phase.

The sustainability of adiabatic evolution significantly depends on the 
optical depth of the shell. Initially, Numerous shock-generated 
electron-positron pairs prevent photons from escaping and regulate
the rest-frame downstream temperature to $T_{\text{h}}^{'}=200\, \text{keV}$.
During the adiabatic expansion, the shell undergoes continuous 
cooling with its temperature following $T^{'}\propto V^{'-\frac{1}{3}}$, 
where $V^{'}$ represents its volume. As the pair density drops sharply 
when the temperature decreases 
below $100 \, \text{keV}$ \citep[e.g.][]{Svensson84},
the optical depth of the breakout shell keeps decreasing until
the pair density becomes negligible at approximate
$T_{\text{th}}^{'}=50 \, \text{keV}$. 
At this point, photons can diffuse out of the breakout shell, 
as its original $\tau_{\text{d}}\lesssim 1$.

For $\gamma_{\text{h}}<\gamma_{\text{h,s}}$, the shell volume 
increases as $V^{'}\propto t/\gamma$ and $V^{'}\propto t$
during and after its acceleration. During the spherical phase,
$V^{'}\propto t^3$. Therefore, when the shell becomes transparent in 
these three phases, the corresponding timescales are
\begin{equation}
t_{\text{h,th}} = 
\begin{cases}
t_0\left(\frac{T_{\text{h}}^{'}}{T_{\text{th}}^{'}}\right)^{3+\sqrt{3}}, & t_{\text{h,th}} \leq t_{\text{f}} \\[1.5ex]
t_0\left(\frac{T_{\text{h}}^{'}}{T_{\text{th}}^{'}}\right)^{3}\gamma_{\text{h}}^{\sqrt{3}}, & t_{\text{f}} < t_{\text{h,th}} \leq t_{\text{h,s}} \\[1.5ex]
t_0\left(\frac{T_{\text{h}}^{'}}{T_{\text{th}}^{'}}\right)\gamma_{\text{h}}^{\frac{\sqrt{3}}{3}}\left(\frac{t_0}{t_{\text{h,s}}}\right)^{\frac{1}{3}}, & t_{\text{h,th}} > t_{\text{h,s}}
\end{cases}
\end{equation}
and the corresponding Lorentz factors of the shell are
\begin{equation}
\gamma_{\text{h,th}} = 
\begin{cases}
\gamma_{\text{h}}\left(\frac{T_{\text{h}}^{'}}{T_{\text{th}}^{'}}\right)^{\sqrt{3}}, & \frac{T_{\text{h}}^{'}}{T_{\text{th}}^{'}} < \gamma_{\text{h}}\\[1.5ex]
\gamma_{\text{f}}, & \gamma_{\text{h}} \leq \frac{T_{\text{h}}^{'}}{T_{\text{th}}^{'}}, \left(\frac{t_0 T_{\text{h}}^{'3}}{t_{\text{h,s}} T_{\text{th}}^{'3}}\right)^{-\frac{\sqrt{3}}{3}} \\[1.5ex]
\gamma_{\text{f}}. & \left(\frac{t_0 T_{\text{h}}^{'3}}{t_{\text{h,s}} T_{\text{th}}^{'3}}\right)^{-\frac{\sqrt{3}}{3}} < \gamma_{\text{h}} \leq \frac{T_{\text{h}}^{'}}{T_{\text{th}}^{'}}
\end{cases}
\end{equation}
For $\gamma_{\text{h}}>\gamma_{\text{h,s}}$, the acceleration persists 
throughout the entire planar phase and ends in the spherical phase. In
such case, Equation (\ref{gacc}) is no longer applicable. However, 
given the rapid expansion of volume, we simplistically assume that 
the breakout shell becomes transparent around the beginning of spherical 
phase. Therefore, the transparency timescale and the Lorentz factor 
of the breakout shell are
\begin{equation} 
\begin{cases}
t_{\text{h,th}} = t_0\left(\frac{T_{\text{h}}^{'}}{T_{\text{th}}^{'}}\right)^{3+\sqrt{3}}, \, \,
\gamma_{\text{h,th}} = \gamma_{\text{h}}\left(\frac{T_{\text{h}}^{'}}{T_{\text{th}}^{'}}\right)^{\sqrt{3}}, 
& t_{\text{h,th}} \leq t_{\text{h,s}} \\[1.5ex]
t_{\text{h,th}} \simeq t_{\text{h,s}}, \, \,
\gamma_{\text{h,th}} \simeq \gamma_{\text{h}}\left(\frac{t_0}{t_{\text{h,s}}}\right)^{\frac{1-\sqrt{3}}{2}}. & t_{\text{h,th}} \gtrsim t_{\text{h,s}}
\end{cases}
\end{equation}

\section{Emission properties of disk-cocoon in stratified AGN disk}
\label{Appendix-Disk-Cocoon}
\setcounter{equation}{0}
When the disk-cocoon vertically expands within a disk 
characterized by a decreasing density profile of the 
form $\rho(z)\propto(z_{\text{c,bre}}-z)^{n}$, where $z_{\text{c,bre}}$
is the shock breakout height of the disk cocoon, the 
velocity of its forefront shock accelerates as 
$\beta_{\text{c,h}}\propto \rho^{-\mu}$, where $\mu\approx 0.19$
\citep{Sakurai60}. Although neither the isothermal nor
the polytropic disk strictly adheres to a power-law decay density
profile, the disk density profile can be locally approximated as
a power-law one by matching the reduction rate at a specific
height, thereby assigning an effective power-law index 
$n_{\text{eff}}$ for different $z_{\text{c,bre}}$ \citep{Grishin21}.
Setting $H_{\text{d}}$ as the lower limit height of the effective 
power-law density profile, the density profile can be expressed as
$\rho_{\text{eff}}(z)=\rho(H_{\text{d}})
[(z_{\text{c,bre}}-z)/(z_{\text{c,bre}}-H_{\text{d}})]^{n_{\text{eff}}}$,
and $n_{\text{eff}}$ can be determined by
$\text{d}\rho_{\text{eff}}(z)/\text{d}z=\text{d}\rho(z)/\text{d}z$.
Therefore, for isothermal disk,
\begin{equation}
n_{\text{eff,iso}}=\frac{z_{\text{c,bre}}-H_{\text{d}}}{H_{\text{d}}},
\end{equation}
and for polytropic disk,
\begin{equation}
n_{\text{eff,poly}}=\frac{6}{5}\frac{z_{\text{c,bre}}-H_{\text{d}}}{H_{\text{d}}}.
\end{equation}
The acceleration occurs at approximately  
$z_{\text{acc,iso}}\sim(3/2\mu)^{1/2}H_{\text{d}}=2.8H_{\text{d}}$ for 
isothermal disk and 
$z_{\text{acc,poly}}\sim(1/6+2\mu/3)^{-\frac{1}{2}}H_{\text{d}}=1.8H_{\text{d}}$ 
for polytropic disk \citep{Grishin21}. When $z_{\text{c,bre}}<z_{\text{acc}}$,
we disregard shock acceleration and model the luminosity and temperature
of the disk-cocoon emission in the same manner as the uniform disk case, 
i.e., Equation (\ref{Lcnacc}), (\ref{TcnaccBB}), and (\ref{TcnaccCom}). 
Conversely, we follow \cite{Nakar10} to investigate the emission evolution.

By estimating the typical velocity of the shocked disk-cocoon shell at 
$z_{\text{acc}}$ as $\beta_{\text{c,d}}\simeq(E_{\text{c}}/M_{\text{c}}c^2)^{\frac{1}{2}}$, 
for the breakout shell, its velocity is accelerated to
$\beta_{\text{c,bre}}\simeq \beta_{\text{c,d}} 
[\rho(z_{\text{c,bre}})/\rho(z_{\text{acc}})]^{-\mu}$, and thereby the breakout
height can be determined using 
$\tau_{\text{d}}(z_{\text{c,bre}})=1/\beta_{\text{c,bre}}$.
The luminosity evolution can be described by \citep{Nakar10, Piro21}
\begin{equation}
L_{\text{c}}(t) = 
\begin{cases}
L_{\text{c,bre}}\exp\left[1-\frac{t_{\text{c,bre}}}{t}\right], & t \leq t_{\text{c,bre}} \\[1.5ex]
L_{\text{c,bre}}\left(\frac{t}{t_{\text{c,bre}}}\right)^{-\frac{4}{3}}, & t_{\text{c,bre}} < t \leq t_{\text{c,pla}} \\[1.5ex]
L_{\text{c,bre}}\left(\frac{t_{\text{c,pla}}}{t_{\text{c,bre}}}\right)^{-\frac{4}{3}}\left(\frac{t}{t_{\text{c,pla}}}\right)^{-\frac{2.28n-2}{3(1.19n+1)}}, & t_{\text{c,pla}} < t \leq t_{\text{c,d}} \\[1.5ex]
L_{\text{c,d}}\exp\left[-\frac{1}{2}\left(\frac{t^2}{t_{\text{c,d}}^2}-1\right)\right], & t > t_{\text{c,d}}
\end{cases}
\end{equation}
where $n=n_{\text{eff,iso}}$ or $n_{\text{eff,poly}}$ for specific disk
vertical density profile;
$L_{\text{c,bre}}$ and $t_{\text{c,bre}}$ are calculated by substituting 
$\beta_{\text{c,bre}}$ into Equation (\ref{Lcbre}) and (\ref{tcbre}); 
$t_{\text{c,d}}$ represents the time when photons initially coupled with the 
accelerated shells have sufficiently diffused out, leading to subsequent 
emission originating from the unaccelerated material, which can be 
calculated by replacing $H_{\text{d}}$ with $z_{\text{acc}}$ in Equation
(\ref{tcdiff}); the corresponding emission luminosity is determined
using Equation (\ref{Lcsph}). Still, for simplicity, we directly connect 
the planar phase with $t<t_{\text{c,pla}}$ and the disk-cocoon cooling phase
with $t>t_{\text{c,d}}$.

For $\beta_{\text{c,bre}}<0.03$, the observed temperature evolves as
\begin{equation}\label{TBBacc}
T_{\text{BB, c}}(t) \propto
\begin{cases}
\exp\left[1-\frac{t_{\text{c,bre}}}{t}\right]^{\frac{1}{4}}, & t \leq t_{\text{c,bre}} \\[1.5ex]
\left(\frac{t}{t_{\text{c,bre}}}\right)^{-\frac{2(9n+5)}{3(17n+9)}}, & t_{\text{c,bre}} < t \leq t_{\text{c,pla}} \\[1.5ex]
\left(\frac{t}{t_{\text{c,pla}}}\right)^{-\frac{18.48n^2+20.69n+6}{(1.19n+1)(22.32n+17)}}, & t_{\text{c,pla}} < t \leq t_{\text{c,d}} \\[1.5ex]
\exp\left[-\frac{1}{2}\left(\frac{t^2}{t_{\text{c,d}}^2}-1\right)\right]^{\frac{4.33n+3}{4(2.79n+2.13)}}\left(\frac{t}{t_{\text{c,d}}}\right)^{-\frac{1.3(n+1)}{2.79n+2.13}}, & t > t_{\text{c,d}}
\end{cases}
\end{equation}
where $T_{\text{c}}(t_{\text{c,bre}}) = T_{\text{BB,cbre}}=
(18\rho\beta_{\text{c,bre}}^2c^2/7a)^{\frac{1}{4}}$, and the evolution during
$t \leq t_{\text{c,d}}$ refers to \cite{Nakar10}. For photons diffusing
out of a series of shells with different mass, velocity, and optical depth,
the observed temperature is determined at the so called color radius, where 
$\eta=1$ \citep{Nakar10}.
Within this color radius framework, we investigate the temperature evolution
during $t > t_{\text{c,d}}$. For a spherically expanding shell, its thermal 
coupling coefficient $\eta \propto T_{\text{BB,sh}}^{\frac{7}{2}} 
\rho_{\text{sh}}^{-2} t_{\text{diff, sh}}^{-1} 
\propto L_{\text{c}}^{\frac{7}{8}} 
m_{\text{sh}}^{-\frac{17}{8}-\frac{1.33n}{2(n+1)}} t^{\frac{7}{2}}$, where
the parameters follow that the thermal equilibrium temperature 
$T_{\text{BB,sh}}\propto L_{\text{c}}^{\frac{1}{4}} 
m_{\text{sh}}^{\frac{1}{4}} r_{\text{sh}}^{-1}$, 
the shell density $\rho_{\text{sh}} \propto m_{\text{sh}} r_{\text{sh}}^{-3}$, 
the diffusion of shell
$t_{\text{diff, sh}} \propto m_{\text{sh}} r_{\text{sh}}^{-1}$, the shell radius
$r_{\text{sh}} \propto \beta_{\text{sh}} t$, and the shell velocity
$\beta_{\text{sh}} \propto m^{-\frac{0.19n}{n+1}}$. Therefore, the properties of 
the color shell evolve following $m_{\text{cl}} 
\propto (L_{\text{c}}^{\frac{7}{8}} t^{\frac{7}{2}})^{\frac{n+1}{2.79n+2.13}}$, 
$\beta_{\text{cl}} \propto (L_{\text{c}}^{\frac{7}{8}} 
t^{\frac{7}{2}})^{-\frac{0.19n}{2.79n+2.13}}$,
and $r_{\text{cl}} \propto (L_{\text{c}}^{\frac{7}{8}} 
t^{\frac{7}{2}})^{-\frac{0.19n}{2.79n+2.13}}t$.
The resulting observed temperature is 
$T_{\text{BB, c}} \propto L_{\text{c}}^{\frac{1}{4}} 
m_{\text{cl}}^{\frac{1}{4}} r_{\text{cl}}^{-1}$, 
i.e., the last row of Equation (\ref{TBBacc}).

For $\beta_{\text{c,bre}}>0.03$, the observed temperature evolves as
\begin{equation}\label{TComacc}
T_{\text{Com, c}}(t) \propto
\begin{cases}
\exp\left[1-\frac{t_{\text{c,bre}}}{t}\right]^{2}\left(\frac{t}{t_{\text{c,bre}}}\right)^{-2}, & t \leq t_{\text{c,bre}} \\[1.5ex]
\left(\frac{t}{t_{\text{c,bre}}}\right)^{-\frac{2}{3}}, & t_{\text{c,bre}} < t \leq t_{\text{c,pla}} \\[1.5ex]
\left(\frac{t}{t_{\text{c,pla}}}\right)^{-\frac{21.26n+1}{3(1.19n+1)}}, & t_{\text{c,pla}} < t \leq t_{\text{c,th1}} \\[1.5ex]
\left(\frac{t}{t_{\text{c,th1}}}\right)^{-\frac{3n+2}{6(1.19n+1)}}, & t_{\text{c,th1}} < t \leq t_{\text{c,th2}} \\[1.5ex]
\left(\frac{t}{t_{\text{c,th2}}}\right)^{-\frac{18.48n^2+20.69n+6}{(1.19n+1)(22.32n+17)}}, & t_{\text{c,th2}} < t \leq t_{\text{c,d}} \\[1.5ex]
\exp\left[-\frac{1}{2}\left(\frac{t^2}{t_{\text{c,d}}^2}-1\right)\right]^{\frac{4.33n+3}{4(2.79n+2.13)}}\left(\frac{t}{t_{\text{c,d}}}\right)^{-\frac{1.3(n+1)}{2.79n+2.13}}, & t > t_{\text{c,d}}
\end{cases}
\end{equation}
where $T_{\text{Com,c}}(t_{\text{c,bre}}) = T_{\text{Com,cbre}}$, 
which is calculated via substituting $\beta_{\text{c,bre}}$ in 
Equation (\ref{eqcom}) to replace $\beta_{\text{h}}$, and the 
evolution during $t \leq t_{\text{c,d}}$ refers to \cite{Nakar10}.
$t_{\text{c,th1}}$ and $t_{\text{c,th2}}$ represent two critical time,
where
\begin{equation}
t_{\text{c,th1}}= t_{\text{c,pla}} \left[\eta_{\text{c,bre}}\left(\frac{t_{\text{c,bre}}}{t_{\text{c,pla}}}\right)^{\frac{1}{6}}\right]^{\frac{3(1.19n+1)}{9.88n+5}},
\end{equation}
\begin{equation}
t_{\text{c,th2}}= t_{\text{c,pla}} \left[\eta_{\text{c,bre}}\left(\frac{t_{\text{c,bre}}}{t_{\text{c,pla}}}\right)^{\frac{1}{6}}\right]^{\frac{6(1.19n+1)}{12.48n+1}}.
\end{equation}
When $t<t_{\text{c,th1}}$, the observed emission temperature is 
out of thermal equilibrium and exhibits Comptonization; when 
$t_{\text{c,th1}}<t<t_{\text{c,th2}}$, emission reaches thermal
equilibrium, and the observed temperature is determined by
the shell from which photons escape; when $t>t_{\text{c,th2}}$, 
photons interact with outer shells during diffusion,
modifying their properties, and the observed temperature is 
determined by the outermost shell that is in thermal equilibrium, 
the evolution reverts to Equation (\ref{TBBacc}). Taking 
Equation (\ref{TComacc}) as the benchmark, the relative magnitudes 
among different timescales alter the specific temperature evolution.
For $t_{\text{c,th1}} < t_{\text{c,d}} \leq t_{\text{c,th2}}$ , since the 
outer shells do not affect the emission, the temperature evolution 
after $t_{\text{c,d}}$ follows that of an isolated shell,
where $T \propto L_{\text{c}}^{\frac{1}{4}} t^{-\frac{1}{2}}$. When 
$t_{\text{c,d}} < t_{\text{c,th1}} $, similarly, after $t_{\text{c,d}}$,
we consider the cooling of an isolated shell. Given that the 
temperature is initially out of thermal equilibrium, it evolves as 
$T \propto t^{-1}$ until thermal equilibrium is reached, after 
which $T \propto L_{\text{c}}^{\frac{1}{4}} t^{-\frac{1}{2}}$. 
Furthermore, if the 
emission approaches thermal equilibrium during the planar phase, 
its temperature will evolve in accordance with Equation (\ref{Tcomp}).

\section{Temperature evolution of Newtonian jet-cocoon}
\label{TcjN}
\setcounter{equation}{0}
To ascertain whether the jet-cocoon emission is in thermal
equilibrium, we calculate the thermal coupling coefficient
of the Newtonian jet-cocoon shells. Before undergoing 
spherical expansion, materials need first escape from the 
initial jet-cocoon cylinder. Assuming a shell is located 
at a distance $d_{\text{cj}}=f_{\text{d}}z_{\text{bre}}$ from
the breakout height, the escaping time can be estimated as 
$t_{\text{esc}}=d_{\text{cj}}/\beta_{\text{cj}}c$. Thus,
for the shell with $\beta_{\text{cj,s}}$, its initial thermal 
coupling coefficient is
\begin{equation}
\eta_{\text{esc,s}}= \frac{aT_{\text{BB,cj0}}^{\frac{7}{2}}}{1.05\times 10^{37}k\rho_{\text{cj,s}}^2t_{\text{esc}}}
=\frac{E_{\text{cj}}^{\frac{7}{8}}V_{\text{cj}}^{\frac{9}{8}}\beta_{\text{cj,s}}c}
{A m_{\text{cj,s}}^2 f_{\text{d}}z_{\text{bre}}},
\end{equation}
where $A=1.05\times 10^{37}k/a^{\frac{1}{8}}$, 
$\rho_{\text{cj,s}}= m_{\text{cj,s}}/V_{\text{cj,s}}$. For heavier 
jet-cocoon material with $\beta_{\text{cj,i}}<\beta_{\text{cj,s}}$,
$T_{\text{BB,cj}}$ is nearly constant due to energy 
density equalization in jet-cocoon, its density is estimated
as $\rho_{\text{cj,i}}/\rho_{\text{cj,s}} = 
m_{\text{cj,i}}d_{\text{cj,s}}/m_{\text{cj,s}}d_{\text{cj,i}}$, 
and its escaping time is approximately
$t_{\text{esc,i}}/t_{\text{esc,s}}=
d_{\text{cj,s}}\beta_{\text{cj,i}}/d_{\text{cj,i}}\beta_{\text{cj,s}}$. 
As $E \propto V_{\text{cj},\beta}$, the spatial distribution of
the jet-cocoon material follows
$\text{d}E/\text{d}\log\beta_{\text{cj}} \propto \text{d}d(>v)/\text{d}
\log\beta_{\text{cj}}$, leading to depth
evolves as $d_{\text{cj,i}}/d_{\text{cj,s}}= (f_{\Gamma \beta} 
z_{\text{bre}}/d_{\text{cj,s}}) \text{log}[\beta_{\text{cj,s}}/\beta_{\text{cj,i}}]+1$.
Consequently, the thermal coupling coefficient 
for heavier shell is given by
\begin{equation}
\eta_{\text{esc,i}}= \eta_{\text{esc,s}} \left(\frac{m_{\text{cj,i}}}{m_{\text{cj,s}}}\right)^{-\frac{2s+3}{s+1}}
\left(\frac{f_{\Gamma \beta} 
z_{\text{bre}}}{d_{\text{cj,s}}}\log\left[\frac{\beta_{\text{cj,s}}}{\beta_{\text{cj,i}}}\right]+1\right).
\end{equation}
During the spherical expansion phase, the shell radius increases
as $r_{\text{cj}}= \beta_{\text{cj}} t$. Consequently, the properties 
of materials outside the luminosity shell evolve according to
$T_{\text{BB,cj}}\propto L_{\text{cj}}^{\frac{1}{4}}
m_{\text{cj}}^{\frac{1}{4}}r_{\text{cj}}^{-1}$,
$\rho_{\text{cj}}\propto m_{\text{cj}} r_{\text{cj}}^{-3}$, and
$t_{\text{cj}}\propto m_{\text{cj}} r_{\text{cj}}^{-1}$, which results 
in the thermal coupling coefficient evolving as
\begin{equation}\label{etacj}
\eta_{\text{cj}}(m_{\text{cj}}>m_{\text{cj,s}})= \eta_{\text{cj,s}}
\left(\frac{m_{\text{cj}}}{m_{\text{cj,s}}}\right)^{-\frac{17s+45}{8(s+1)}}
\left(\frac{t}{t_{\text{cj,s}}}\right)^{\frac{7(s+1)}{2(s+2)}},
\end{equation}
where
\begin{equation}
\eta_{\text{cj,s}}=\eta_{\text{esc,s}}\left(\frac{t_{\text{cj,s}}}{t_{\text{esc,s}}}\right)^{\frac{3}{2}},
\end{equation}
is the thermal coupling coefficient of the shell with 
$\beta_{\text{cj,s}}$ at $t_{\text{cj,s}}$, derived from
the adiabatic expansion of the shell from $t_{\text{esc,s}}$
to $t_{\text{cj,s}}$. Since the thermal coupling becomes weaker 
during the spherical phase, $\eta_{\text{esc,s}}$ governs
the temperature properties of the jet-cocoon emission.

When $\eta_{\text{esc,s}}\leqslant 1$, the emission is in
thermal equilibrium, and its temperature evolves as
\begin{equation}\label{TBBcj}
T_{\text{BB,cj}}(t) \propto
\begin{cases}
\left(\frac{t}{t_{\text{cj,s}}}\right)^{-\frac{2(5s^2+27s+50)}{(s+2)(17s+45)}}, & t_{\text{cj,s}} < t \leq t_{\text{th,ph}} \\[1.5ex]
\left(\frac{t}{t_{\text{th,ph}}}\right)^{-\frac{s^2+5s+8}{2(s+2)(s+3)}}, & t_{\text{th,ph}} < t \leq t_{\text{sph,end}} \\[1.5ex]
\exp\left[-\frac{1}{2}\left(\frac{t^2}{t_{\text{sph,end}}^2}-1\right)\right]^{\frac{1}{4}}\left(\frac{t}{t_{\text{sph,end}}}\right)^{-\frac{s+1}{2(s+3)}}, & t > t_{\text{sph,end}}
\end{cases}
\end{equation}
where $T_{\text{BB,cj}}(t_{\text{cj,s}})=T_{\text{BB,cjs}}$, and
$T_{\text{BB,cjs}}$ is the thermal equilibrium temperature 
at $t_{\text{cj,s}}$, which is given by
\begin{equation}
T_{\text{BB,cjs}}=\left(\frac{L_{\text{cj,s}}}{a c \beta_{\text{cj,s}}r_{\text{cj,s}}^2}\right)^{\frac{1}{4}}.
\end{equation} 
By setting $\eta_{\text{cj}}=1$ in Equation (\ref{etacj}), the 
properties of the color shell, which determines the observed 
temperature \citep{Nakar10}, can be achieved. The mass 
outside this shell is 
\begin{equation}
m_{\text{cj,cl}}= m_{\text{cj,s}} \eta_{\text{cj,s}}^{\frac{8(s+1)}{17s+45}}
\left(\frac{t}{t_{\text{cj,s}}}\right)^{\frac{28(s+1)^2}{(s+2)(17s+45)}},
\end{equation}
and as $m(>v)\propto v^{-(s+1)}$, the color radius is 
given by
\begin{equation}
r_{\text{cj,cl}}=r_{\text{cj,s}} \eta_{\text{cj,s}}^{-\frac{8}{17s+45}}   
\left(\frac{t}{t_{\text{cj,s}}}\right)^{\frac{17s^2+51s+62}{(s+2)(17s+45)}}.
\end{equation}
Nevertheless, when photons diffuse out of the jet-cocoon
photosphere, material's influence on the emission becomes 
negligible, leading to the failure of the color shell, 
and we assume that the observed temperature is determined 
by the photosphere radius, which is at 
$\kappa m/4\pi r_{\text{cj,ph}}^2=1$, i.e.,
\begin{equation}
r_{\text{cj,ph}}=r_{\text{cj,s}} \beta_{\text{cj,s}}^{-\frac{1}{s+3}} 
\left(\frac{t}{t_{\text{cj,s}}}\right)^{\frac{s+1}{s+3}}.
\end{equation}
The transition occurs at $r_{\text{cj,cl}}=r_{\text{cj,ph}}$,
where the corresponding time is
\begin{equation}
t_{\text{th,ph}}= t_{\text{s, cj}} \beta_{\text{cj,s}}^{-\frac{(s+2)(17s+45)}{6s^2+46s+96}} 
\eta_{\text{cj,s}}^{\frac{4(s+2)(s+3)}{3s^2+23s+48}}.
\end{equation}
When $t<t_{\text{th,ph}}$, the observed temperature evolves 
following $T_{\text{BB,cj}} \propto L_{\text{cj}}^{\frac{1}{4}} 
m_{\text{cj,cl}}^{\frac{1}{4}} r_{\text{cj,cl}}^{-1}$, thereby
yielding the first row of Equation (\ref{TBBcj}). 
Similarly, when $t>t_{\text{th,ph}}$, the observed 
temperature follows $T_{\text{BB,cj}} \propto 
L_{\text{cj}}^{\frac{1}{4}} r_{\text{cj,ph}}^{-1}$, leading to
the last two rows of Equation (\ref{TBBcj}).

For the jet-cocoon with $\eta_{\text{esc,s}}>1$, its temperature 
is initially out of thermal equilibrium. The Compton modified 
temperature can be calculated using 
$T_{\text{esc}} \xi(T_{\text{esc}})^2 =T_{\text{BB,cj0}}
\eta_{\text{esc}}(T_{\text{BB,cj0}})^2$ \citep{Nakar10}, 
where the Comptonization correction factor is
\begin{equation}
\xi(T) \approx \max\left\{1,\frac{1}{2}\ln\left(y_{\text{max}}\right)
\left[1.6+\ln\left(y_{\text{max}}\right)\right]\right\},
\end{equation}
with $y_{\text{max}}=3~(\rho/10^{-9}\text{\,g\,cm}^{-3})^{-\frac{1}{2}}
(T/100\text{\,eV})^{\frac{9}{4}}$.
After adiabatic cooling, the observed temperature at 
$t_{\text{cj,s}}$ is
\begin{equation}
T_{\text{Com,cjs}}\simeq T_{\text{esc,s}}\frac{V_{\text{cj,s}}^{\frac{1}{3}}}{r_{\text{cj,s}}}.
\end{equation}
The temperature evolution is given by
\begin{equation}\label{Tcomcj}
T_{\text{Com,cj}}(t) \propto
\begin{cases}
\left(\frac{t}{t_{\text{cj,s}}}\right)^{-\frac{9s+12}{s+2}}, & t_{\text{cj,s}} < t \leq t_{\text{cj,th1}} \\[1.5ex]
\left(\frac{t}{t_{\text{cj,th1}}}\right)^{-\frac{s+1}{2(s+2)}}, & t_{\text{cj,th1}} < t \leq t_{\text{cj,th2}} \\[1.5ex]
\left(\frac{t}{t_{\text{cj,th2}}}\right)^{-\frac{2(5s^{2}+27s+50)}{(s+2)(17s+45)}}, & t_{\text{cj,th2}} < t \leq t_{\text{th,ph}} \\[1.5ex]
\left(\frac{t}{t_{\text{th,ph}}}\right)^{-\frac{s^2+5s+8}{2(s+2)(s+3)}}, & t_{\text{th,ph}} < t \leq t_{\text{sph,end}} \\[1.5ex]
\exp\left[-\frac{1}{2}\left(\frac{t^2}{t_{\text{sph,end}}^2}-1\right)\right]^{0.25}\left(\frac{t}{t_{\text{sph,end}}}\right)^{-\frac{s+1}{2(s+3)}}, & t > t_{\text{sph,end}}
\end{cases}
\end{equation}
where $T_{\text{cj}}(t_{\text{cj,s}})=T_{\text{Com,cjs}}$. Similar 
to the disk-cocoon case, two critical times 
$t_{\text{cj,th1}}\equiv t(\eta_{\text{esc}}=1)$ and 
$t_{\text{cj,th2}}\equiv t(\eta_{\text{cj}}=1)$ calculating at 
the luminosity shell are
\begin{equation}
t_{\text{cj,th1}}\simeq t_{\text{cj,s}} \eta_{\text{esc,s}}^{\frac{s+2}{2(2s+3)}},
\end{equation}
\begin{equation}
t_{\text{cj,th2}}= t_{\text{cj,s}} \eta_{\text{cj,s}}^{\frac{4(s+2)}{3s+31}}.
\end{equation}
During $t_{\text{cj,s}} < t \leq t_{\text{cj,th1}}$, the observed temperature 
is determined by the luminosity shell, given by 
$T_{\text{Com,cj}}\simeq\eta_{\text{esc}}^2T_{\text{BB,0}} 
(V_{\text{cj},\beta}^{\frac{1}{3}}/r_{\text{cj}})
\propto t^{-\frac{9s+12}{s+2}}$.
The adiabatic cooling of the luminosity shell 
governs the temperature evolution when
$t_{\text{cj,th1}} < t \leq t_{\text{cj,th2}}$, 
resulting in
$T_{\text{Com,cj}}\propto L_{\text{cj}}^{\frac{1}{4}} 
m_{\text{cj}}^{\frac{1}{4}} r_{\text{cj}}^{-1}
\propto t^{-\frac{s+1}{2(s+2)}}$. After
$t>t_{\text{cj,th2}}$, the temperature evolution
follows Equation (\ref{TBBcj}).

Furthermore, if the Compton modified temperature 
is exceedingly high, the newly generated pairs
would suppress the temperature to approximately 
$200\, \text{keV}$ when materials leave 
the jet-cocoon cylinder. During the adiabatic
spherical expansion, the pair effect becomes
negligible in trapping photons when the 
temperature decreases to $\sim 50\, \text{keV}$
\citep{Nakar12}, at a time given by
\begin{equation}
t_{\text{cj,np}} (\beta_{\text{cj}}) \simeq
\frac{V_{\text{cj},\beta}^{\frac{1}{3}}}{\beta_{\text{cj}}c}
\frac{200\, \text{keV}}{50\, \text{keV}}
=\frac{4 V_{\text{cj},\beta}^{\frac{1}{3}}}{\beta_{\text{cj,}}c}.
\end{equation}
For $t_{\text{cj,np}}(\beta_{\text{cj,s}})<t_{\text{cj,s}}$, photons
escape and produce emission at diffusion radius 
$r_{\text{cj,s}}$ with the observed temperature given by
\begin{equation}
T_{\text{pair,cjs}}= 200\, \text{keV} 
\frac{V_{\text{cj,s}}^{\frac{1}{3}}}{r_{\text{cj,s}}}.
\end{equation}
As the initial temperatures of the luminosity 
shells remain nearly constant due to the 
regulation of pairs, the observed temperature
evolves following $T_{\text{pair,cj}}\propto 
r_{\text{cj}}^{-1}$ of the luminosity shell, i.e.,
\begin{equation}
T_{\text{pair,cj}}(t)=T_{\text{pair,cjs}} 
\left(\frac{t}{t_{\text{cj,s}}}\right)^{-\frac{s}{s+2}}, 
\quad t_{\text{cj,s}} < t \leq t_{\text{esp,np}}
\end{equation}
where $t_{\text{esp,np}}$ is the time when the 
initial Comptonized temperature $T_{\text{esc}}$ 
decreases to $50\, \text{keV}$ with few pair
production. By neglecting the evolution of $\xi$,
$T_{\text{esc}}\simeq T_{\text{esc,s}} 
(\eta_{\text{esc}}/\eta_{\text{esc,s}})^2$, and
thereby 
\begin{equation}
t_{\text{esp,np}}\simeq t_{\text{cj,s}} \left(\frac{T_{\text{esc,s}}}
{50\, \text{keV}}\right)^{\frac{s+2}{4(2s+3)}}.
\end{equation}
When $t>t_{\text{esp,np}}$, the observed temperature
follows Equation (\ref{Tcomcj}) by substituting 
$t_{\text{esp,np}}$ for $t_{\text{cj,s}}$. However, 
for $t_{\text{cj,np}}(\beta_{\text{cj,s}})>t_{\text{cj,s}}$, 
the luminosity shell must propagate 
to a radius larger than 
$r_{\text{cj,s}}$ to produce emission with a temperature
of $50\, \text{keV}$. The condition continues until the
velocity of the luminosity shell satisfies
$t_{\text{cj,np}}=t_{\text{cj}}$, which is given by
\begin{equation}
\beta_{\text{cj,50keV}}=\frac{\kappa f_{\Gamma \beta} E_{\text{cj}}}
{32\pi c^2 V_{\text{cj},\beta}^{\frac{2}{3}}},
\end{equation}
and the corresponding time is
\begin{equation}
t_{\text{cj,50keV}}=\frac{128\pi c V_{\text{cj},\beta}}
{\kappa f_{\Gamma \beta} E_{\text{cj}}}.
\end{equation}
Thus, the observed temperature evolves as
\begin{equation}
T_{\text{pair,cj}}(t) =
\begin{cases}
50\, \text{keV}, & t_{\text{cj,np}}(\beta_{\text{cj,s}}) < t \leq t_{\text{cj,50keV}} \\[1.5ex]
50\, \text{keV}\left(\frac{t}{t_{\text{cj,50keV}}}\right)^{-\frac{s}{s+2}}. 
& t_{\text{cj,50keV}} < t \leq t_{\text{esp,np}}
\end{cases}
\end{equation}
Still, when $t>t_{\text{esp,np}}$, the observed temperature
follows Equation (\ref{Tcomcj}) by substituting 
$t_{\text{esp,np}}$ for $t_{\text{cj,s}}$.
\end{appendix}

\end{document}